\documentclass[11pt,a4paper,english]{article}
\pdfoutput=1
\usepackage[T1]{fontenc}
\usepackage[utf8]{inputenc}
\usepackage{color}
\usepackage{amsbsy}
\usepackage{amstext}
\usepackage{amssymb}
\usepackage{graphicx}

\makeatletter

\pdfpageheight\paperheight
\pdfpagewidth\paperwidth

\usepackage{jcappub}
\usepackage{amsmath}
\usepackage{esint}
\pdfoutput=1

\renewcommand\[{\begin{equation}}
\renewcommand\]{\end{equation}}

\author[a,b]{Felipe O. Franco,}
\author[a]{Thiago S. Pereira}

\affiliation[a]{Departamento de Física, Universidade Estadual 
de Londrina, 86057-970, Londrina PR, Brazil}
\affiliation[b]{Département de Physique Théorique and Center for Astroparticle Physics 
(CAP), University of Geneva, 24 quai Ernest Ansermet, CH-1211 Geneva, Switzerland}

\emailAdd{Felipe.Oliveira@unige.ch}
\emailAdd{tspereira@uel.br}

\abstract{Besides expanding anisotropically, the universe can 
also be anisotropic at the level of its (spatial) curvature. In particular,
models with anisotropic curvature and isotropic expansion leads
both to a $\Lambda$CDM-like phenomenology and to an isotropic and homogeneous CMB
at the background level. Thus, they offer an interesting and viable example 
where the cosmological principle does not follow from the isotropy of  
observational data. In this paper we extract the linear dynamics of tensor perturbations 
in two classes of cosmologies with anisotropic spatial curvature. Two difficulties 
arise in comparison to the same computation in isotropic 
cosmologies. First, the two tensor polarizations do not behave as a spin-2 field, 
but rather as the spin-0 and spin-1 irreducible components
of a symmetric, traceless and transverse tensor field, each with its own dynamics.
Second, because metric perturbations are algebraically coupled, one cannot ignore
scalar and vector modes and focus just on tensors --- even if one is 
only interested in the latter --- under the penalty of obtaining the 
wrong equations of motion. We illustrate our results by finding analytical
solutions and evaluating the power-spectra of tensor polarizations in a radiation 
dominated universe. We conclude with some comments on how these models could be 
constrained with future experiments on CMB polarization.}


\makeatother

\usepackage{babel}
\begin{document}

\title{Tensor Perturbations in Anisotropically Curved Cosmologies}

\maketitle
\global\long\def\ph{\phantom{\mu}}

\section{Introduction}

The recent detection of gravitational waves through a binary system
of black holes by the LIGO and VIRGO collaborations \cite{PhysRevLett.116.061102,Abbott:2017vtc,Abbott:2016nmj}
is arguably a landmark in the history of science. In addition to confirming
General Relativity as a full-fledged theory of gravitational interactions,
it represents the birth of a new observational era to physicists and
astronomers, from which unfathomable new features of the universe
might arise.

From the perspective of cosmology, the detection of gravitational
waves from astrophysical processes renews our hope that primordial
gravitational waves might also be detected in the coming future. Such
discovery would have a huge impact to cosmology, since primordial
gravitational waves are believed to be the fingerprint of an early
(and yet not fully understood) inflationary stage of the universe.
While the path leading to this discovery is still being worked upon,
there are a number of cosmological phenomena closely linked to the
physics of primordial gravitational waves that can help us to indirectly
measure them. For instance, the dynamics of primordial gravitational
waves is supposed to leave an imprint both in the temperature spectrum
of the Cosmic Microwave Background (CMB) radiation — which was measured
with exquisite accuracy by WMAP \cite{Bennett:2012zja} and Planck
\cite{Ade:2015xua} collaborations — and in the polarization spectrum
of $B$-modes of CMB, which is the main goal of current and future
CMB missions \cite{Hanson:2013hsb,Ade:2014xna,Ade:2014afa}. For this
reason, it is an important task to investigate the dynamics of gravitational
waves under given cosmological hypotheses and compare them with cosmological
data.

Indeed, one of the central hypothesis believed to be backed up by
data like galaxy distribution from the Sloan Digital Sky Survey \cite{0004-637X-624-1-54}
and CMB temperature fluctuations by the Planck satellite \cite{Ade:2015hxq},
is that around 100 Mpc and above, the average spatial distribution
of matter in the universe is isotropic. Such spherical symmetry around
us together with the Copernican principle — according to which our
position in the universe is not special — constitutes a hallmark of
modern cosmology known as the Cosmological Principle. Based on this
principle, cosmology has had an incredible advance in the last decades,
from both theoretical and observational points of view, culminating
in a very successful concordance cosmological model, the $\Lambda$CDM
model.

Although very successful in describing the large-scale structure of
the universe — even though several problems still persist \cite{Bull:2015stt}
— one should not lose sight of the fact that the $\Lambda$CDM depends
on untested assumptions about the symmetries of matter distribution
at cosmological scales. Thus, such assumptions should not prevent
us from considering alternative cosmological models which are compatible
with the data. In fact, in the past few years several authors have
considered the possibility of giving up the Cosmological Principle
in a way or another. Some examples of these are large void models
which attempt to fit the spectrum of CMB assuming that we would be
located near the center of a spherically symmetrical universe \cite{GarciaBellido:2008nz,Nadathur:2010zm,Biswas:2010xm,Yoo:2010qy},
and spatially homogeneous but anisotropic cosmological models which
are justified either as an explanation for the statistical anomalies
of the CMB \cite{Campanelli:2006vb,Gumrukcuoglu:2006xj,Pontzen:2007ii,Rodrigues:2007ny}
or as a mean of constraining the impact of spatial anisotropy on CMB
data \cite{martinez1995delta,Maartens:1994qq,Maartens:1995hh,Pitrou:2008gk}.
The interest in inhomogeneous and anisotropic models has in fact a
richer history prior to modern CMB data — see for example \cite{Barrow:1985tda,Barrow:1997mj,Barrow:1997sy,Barrow:1998ih,Barrow:2001pi}.

Here we contribute to this task by investigating a class of homogeneous
but spatially anisotropic models such that their time-constant hypersurfaces
have a preferred direction, but which nonetheless preserve the observed
isotropy of CMB at first order (i.e., without including perturbations).
This is possible because the anisotropy of these spacetimes results
from the curvature of the spatial sections, and not from the kinematics
of expansion \cite{Pereira:2012ma,Pereira:2015pxa}. 
From a fundamental perspective, 
such as in string inspired models of the early universe, one can argue that the early 
universe’s dimensions were initially small and compact~\cite{Brandenberger:1988aj}. 
In this scenario, it is plausible that a process of decompactification of 
our three spatial dimensions acts anisotropically (via, e.g., anisotropic 
bubble nucleation~\cite{BlancoPillado:2010uw}), producing a universe with different topologies 
along different spatial dimensions~\cite{Adamek:2010sg,Graham:2010hh}.

Specifically, we focus on two particular solutions of Einsteins equations with these
properties, namely, Bianchi type III (BIII) and Kantowski-Sachs (KS)
metrics. In 1993, Mimoso and Crawford \cite{Mimoso:1993ym} showed
that these metrics admit a geodesic, irrotational and isotropic (or
shear-free) expansion provided that the anisotropic stress-tensor
of the cosmological budget is in direct proportion to the electric
part of the Weyl tensor. The interplay between the energy-momentum
tensor and the shear-free condition was further explored in \cite{Coley:1994yt,McManus:1994ys}.
An analogous result was also found by Carneiro and Marugán \cite{Carneiro:2001fz,cmm}
who deployed a clever use of an anisotropic scalar field to balance
the anisotropy of the spatial curvature. Under this condition the
scale factor has the same dynamics of a spatially curved Friedmann-Lemaître-Robertson-Walker
(FLRW) universe and the metric can be brought to a conformally static
form. It then follows that the electromagnetic radiation, e.g. CMB,
will be isotropic, in accordance with basic observations \cite{Ehlers:1966ad,Clarkson:1999yj},
even though the geometry is fundamentally anisotropic. Thus, these
models offer an important lesson: contrarily to common belief, the
symmetries of cosmological data do not necessarily imply the symmetries
of the cosmic geometry \cite{Coley:1994yt}.

The background phenomenology of BIII and KS shear-free cosmologies
was originally explored in refs. \cite{Mimoso:1993ym,Carneiro:2001fz}.
In \cite{Koivisto:2010dr}, the impact of a directional spatial curvature
on the distribution of type Ia supernovae was also investigated. The
linear and gauge-invariant perturbation theory in such models was
also shown to be viable in \cite{Pereira:2012ma}, and in several
ways parallels that of standard perturbation theory in FLRW spacetimes,
though the expected observational signatures are of course different
\cite{Pereira:2015pxa}. In this work, motivated by the recent detection
of gravitational waves, we focus on the linear dynamics of tensor
perturbations in these models. We thus begin \S\ref{sec:back-model}
by briefly reviewing the geometrical and dynamical aspects of geometries
with anisotropic curvature. Since our goal is to find the dynamics
of tensor perturbations in spacetimes with, as we shall see, a residual
rotational symmetry, and then later contrast our results with tensor
perturbations in FLRW spacetimes, we present in \S\ref{sec:spacetime-splitting}
a systematic description of the irreducible decomposition of symmetric
tensors in the 1+3 and 1+2+1 spacetime splittings, together with a
dictionary to go from one to the other. These steps will allow us
to pinpoint the genuine gauge-invariant tensor degrees of freedom
in a spacetime with anisotropic curvature in \S\ref{sec:Gravitational-waves}.
Next, we use linear perturbation theory to find the equations of motion
of the tensorial degrees of freedom in \S\ref{sec:Dynamical-equations-and},
after which we present some simple applications. As we shall see,
and differently from what happens with perturbation theory in FLRW
spacetimes, this task requires that we keep track of all metric degrees
of freedom — including those not related to gravitational waves —
since algebraic couplings between different modes cannot be ignored,
under penalty of leading to the wrong dynamics of gravitational waves.
We finally conclude and give some perspectives of further developments
in \S\ref{sec:Conclusions}.

Throughout this work, dots will represent derivatives along time,
whereas a prime is reserved for derivatives along the anisotropic
spatial direction. We also adopt units such that $c=1=8\pi G$, and
spacetime signature $\left(-,+,+,+\right)$. 

\section{Background model\label{sec:back-model}}

In this work we are interested in a class of homogeneous and conformally
static geometries codified by the following line element: 
\begin{align}
{\rm d}s^{2} & =a^{2}(\eta)\left[-{\rm d}\eta^{2}+\gamma_{ij}\left(\mathbf{x}\right){\rm d}x^{i}{\rm d}x^{j}\right],\label{bckgd-metric}
\end{align}
where $\eta$ is the usual conformal time parameter and $\gamma_{ij}(\mathbf{x})$
is the spatial metric. Because $\partial_{\eta}\gamma_{ij}(\mathbf{x})=0$,
these models have no geometrical shear, and the expansion is isotropic.
However, we can still have spatial anisotropies if $\gamma_{ij}$
describes manifolds whose (spatial) curvature is direction-dependent.
Since one-dimensional manifolds are trivially flat, the simplest examples
one can think of (i.e., without invoking non-trivial topologies) are
given by metrics on spaces of the form ${\cal M}^{2}\times\mathbb{R}$,
where ${\cal M}^{2}$ is a two-dimensional maximally symmetric space.
Adopting a coordinate system where the real line coincides with the
$z$-axis, we have 
\[
\gamma_{ij}{\rm d}x^{i}{\rm d}x^{j}=\mathcal{S}_{ab}(x^{c}){\rm d}x^{a}{\rm d}x^{b}+{\rm d}z^{2}\,,
\]
where $\{a,b,c\}$ run from 1 to 2 and ${\cal S}_{ab}$ is the metric
on ${\cal M}^{2}$. Because of the residual symmetry of these spaces,
it is convenient to employ cylindrical coordinates so that
\begin{equation}
{\rm d}s^{2}=a^{2}(\eta)\left[-{\rm d}\eta^{2}+{\rm d}\rho^{2}+S_{\kappa}^{2}(\rho){\rm d}\varphi^{2}+{\rm d}z^{2}\right]\,,\label{eq:bckgd-metric cylindrical}
\end{equation}
where the function $S_{\kappa}(\rho)$ is defined as 
\begin{equation}
S_{\kappa}(\rho)=\frac{1}{\sqrt{\kappa}}\sin(\sqrt{\kappa}\rho)\,.\label{S_k}
\end{equation}
Note that, for $\kappa<0$, this parameterization gives $S_{\kappa}=\sinh(\sqrt{|\kappa|}\rho)/\sqrt{|\kappa|}\rho$,
as it should. We adopt a convention where the scale factor is dimensionless
and comoving coordinates have dimension of length. Thus, the curvature
parameter $\kappa$ has units of $({\rm length})^{-2}$, being either
negative, positive or zero. Topologically, this means that ${\cal M}^{2}$
corresponds to either the pseudo-sphere $\mathbb{H}^{2}$ ($\kappa<0$),
the two-sphere $\mathbb{S}^{2}$ $(\kappa>0)$ or the plane $\mathbb{R}^{2}$
($\kappa=0$). In cosmology, this leads to the solutions of Bianchi
type-III (BIII), Kantowski-Sachs (KS) and flat Friedmann-Lemaître-Robertson-Walker
(FLRW) universes, respectively. The flat case is included only for
consistency, since it allows us to verify the validity of our results
in the limit $|\kappa|\ll1$. Thus, in what follows we treat $\kappa$
as a free parameter.

\subsection{Shear-free dynamics}

The family of spacetimes described by (\ref{eq:bckgd-metric cylindrical})
are usually known as spacetimes with anisotropic curvature, or shear-free
spacetimes. These adjectives stem from the already mentioned fact
that the spacetime expansion is isotropic even though the metric is
genuinely anisotropic. But since anisotropic models cannot simultaneously
exhibit shear-free expansion and a perfect fluid matter content \cite{Mimoso:1993ym},
shear-free models can only be realized at the expense of an imperfect
energy-momentum tensor of the form
\begin{equation}
T_{\mu\nu}=(\rho+p)u_{\mu}u_{\nu}+pg_{\mu\nu}+\pi_{\mu\nu}\,,\label{eq:tmunu-imperfect-fluid}
\end{equation}
with the condition $u^{\mu}\pi_{\mu\nu}=0=\pi_{\ph\mu}^{\mu}$. In
the coordinate system (\ref{eq:bckgd-metric cylindrical}), Einstein
background equations are
\begin{align*}
\frac{3\mathcal{H}^{2}}{a^{2}}+\frac{\kappa}{a^{2}} & =\rho_{f}+\rho\,,\\
\frac{{\cal H}^{2}}{a^{2}}+2\frac{\dot{{\cal H}}}{a^{2}} & =-p_{f}-p-\pi_{\ph1}^{1}\,,\\
\frac{{\cal H}^{2}}{a^{2}}+2\frac{\dot{{\cal H}}}{a^{2}}+\frac{\kappa}{a^{2}} & =-p_{f}-p-\pi_{\ph3}^{3}\,,
\end{align*}
where $\rho_{f}$ and $p_{f}$ are the total energy density and pressure
of additional perfect fluids. Note that the residual symmetry of the
metric implies that $\pi_{\ph1}^{1}=\pi_{\ph2}^{2}$, while the trace-free
condition on $\pi_{\mu\nu}$ gives $\pi_{\ph3}^{3}=-2\pi_{\ph1}^{1}$.
Consistency between the last two equations further requires that
\begin{equation}
\pi_{\ph1}^{1}=\frac{\kappa}{3a^{2}}\,.\label{eq:shear-free-cond1}
\end{equation}
This is known as the shear-free condition, and it follows from our
imposition of metric (\ref{eq:bckgd-metric cylindrical}) as a solution
of Einstein equations\footnote{This is equivalent to demanding that $\pi_{\mu\nu}=2E_{\mu\nu}$,
where $E_{\mu\nu}$ is the electric part of the Weyl tensor. See ref.
\cite{Mimoso:1993ym}}. From the phenomenological point of view, the simplest model implementing
(\ref{eq:shear-free-cond1}) is perhaps that of a massless scalar
field with Lagrangian
\[
{\cal L}_{\phi}=-\frac{\alpha}{2}\partial_{\mu}\phi\partial^{\mu}\phi\,,
\]
where $\alpha$ is a constant. Written in the form (\ref{eq:tmunu-imperfect-fluid}),
the energy-momentum tensor of the field $\phi$ gives
\[
\rho=\frac{\alpha}{2}\partial_{\lambda}\phi\partial^{\lambda}\phi\,,\quad p=-\frac{1}{3}\rho\,,\quad\pi_{\ph\nu}^{\mu}=\alpha\partial^{\mu}\phi\partial_{\nu}\phi-\frac{2}{3}\left(\delta_{\nu}^{\mu}+u^{\mu}u_{\nu}\right)\rho\,.
\]
The condition $u_{\mu}\pi_{\ph\nu}^{\mu}=0$ implies that $\partial_{\eta}\phi=0$,
which tell us that $\phi$ has no dynamics. It is easy to verify that
$\phi=z$ satisfies this condition as well as the wave equation $\square\phi=0$
\cite{Carneiro:2001fz}. This implies that
\begin{equation}
\pi_{\ph1}^{1}=-\frac{2}{3}\rho=2p=-\frac{\alpha}{3a^{2}}\,,\label{eq:phi-energy-density}
\end{equation}
which, when compared to (\ref{eq:shear-free-cond1}), gives $\alpha=-\kappa$.
Note that, while $\phi$ is inhomogeneous, its energy density, pressure
and stress are not, so that the homogeneity of the background is preserved.
Such field could arise, for example, in a field-theoretic description
of a solid inflating the universe \cite{Endlich:2012pz,Bartolo:2013msa}.
The phenomenology leading to (\ref{eq:shear-free-cond1}) can also
be obtained with a three-form (Kalb-Ramon) field \cite{Koivisto:2010dr}
and, in general, shear-free solutions can be obtained with $p$-form
gauge fields \cite{Throsrud:2017}. Luckily for us, the precise form
of the field leading to a shear-free expansion will not affect the
dynamics of tensor perturbations, as we will see. The effective background
equations then become
\begin{equation}
\begin{split}{\cal H}^{2} & =\frac{\rho_{f}a^{2}}{3}-\frac{\kappa}{2}\,,\\
\frac{\ddot{a}}{a} & =\frac{1}{6}(\rho_{f}-3p_{f})a^{2}-\frac{\kappa}{2}\,.
\end{split}
\label{eq:eff-flrw-eqs}
\end{equation}
Although not written in a standard from, these are essentially Friedmann
equations in a background with curvature $\kappa/2$ (see, e.g., \cite{Mukhanov:2005sc}).
Since $\kappa$ is a free parameter, it is convenient to introduce
a curvature radius $R_{c}$ through
\begin{equation}
\left|\kappa\right|\equiv\frac{1}{R_{c}^{2}}=2H_{0}^{2}\left|\Omega_{\kappa0}\right|=\frac{2}{L_{c}^{2}}\,,\label{eq:Rc}
\end{equation}
where $H_{0}$ is today's Hubble parameter in physical time and $L_{c}=H_{0}^{-1}\left|\Omega_{\kappa0}\right|^{-1/2}$
is the curvature radius of FLRW metrics. 

\section{Spacetime splittings\label{sec:spacetime-splitting}}

Our main goal is to derive the dynamics of tensor perturbations evolving
in the metric (\ref{eq:bckgd-metric cylindrical}), with the background
evolution being given by (\ref{eq:eff-flrw-eqs}). However, because
the background space is not maximally symmetric, the very definition
of tensor perturbations needs to be revisited. We thus start this
section by briefly recalling the definitions and properties of the
1+3 spacetime splitting, followed by a review of its less notorious
cousin, the 1+2+1 splitting. Such splittings will be combined with
(versions of) the Scalar-Vector-Tensor (SVT) decomposition in the
next section to arrive at a proper definition of gravitational waves
in spacetimes with anisotropic spatial curvature. We stress that our
goal is not to conduct a fully covariant characterization of gravitational
wave dynamics — to which treatment we point the reader to refs. \cite{Hawking:1966qi,Hogan-Ellis,Dunsby:1998hd,Clarkson:2002jz,Osano:2006ew}.
Rather, we use a covariant approach only to identify the ``pure''
tensor modes in spacetimes with anisotropic curvature, after which
we will adopt a more direct coordinate-based approach to find the
dynamics of gravitational waves.

Since the introduction of a new algebraic splitting brings with it
a zoo of new objects, each of which requires a special symbol, it
is important to introduce a notation that avoids the proliferation
of overbars, tildes and the like. \emph{Thus, we adopt the convention
where spacetime (i.e., non-projected) tensors will be represented
by lower case (Latin or Greek) letters; e.g, $a^{\mu}$, $g_{\mu\nu}$
and so on are spacetime quantities. In any 1+3 irreducible decomposition,
each term will be represented by a capital Latin letter. Thus, objects
like $V_{\mu}$, $T_{\mu\nu}$ and so on are by definition orthogonal
to the observer's four velocity $u^{\mu}$, whereas $A$, $B$ and
so on denote scalars ``along'' $u^{\mu}$. Likewise, in any 1+2+1
irreducible decomposition, each term will be represented by calligraphic
capital Latin letter. Hence, terms like ${\cal V}_{\mu}$, ${\cal T}_{\mu\nu}$,
and so on are orthogonal to both $u^{\mu}$ and $N^{\mu}$ (the curvature
privileged direction); likewise, ${\cal A}$, ${\cal B}$ and so on
are scalars ``along'' either $u^{\mu}$ or $N^{\mu}$. The only
exception to this rule is the fully-projected spatial metric $\gamma_{\mu\nu}$,
which we keep as it is to convey to the standard literature. }Moreover,
we shall refer to vectors like $a_{\mu}$, $V_{\mu}$, and ${\cal V}_{\mu}$
as 4-, 3- and 2-vector, respectively. Occasionally, we shall also
refer (somewhat misleadingly but hopefully useful) to tensors like
$g_{\mu\nu}$, $T_{\mu\nu}$ and ${\cal T}_{\mu\nu}$ as 4-, 3- and
2-tensor, respectively.

\subsection{1+3 spacetime splitting\label{subsec:1+3}}

In FLRW cosmologies with perfect-fluid matter content, the geodesic
flow of matter naturally defines a congruence of timelike curves $x^{\mu}\left(\tau\right)$
such that freely falling observers with proper time $\tau$ are fully
characterized by their four-velocity vector $u^{\mu}=dx^{\mu}/d\tau$,
normalized so that $u^{\mu}u_{\mu}=-1$. Such field of vectors naturally
define two projection tensors
\begin{equation}
t_{\ph\nu}^{\mu}\equiv-u^{\mu}u_{\nu}\,,\qquad\gamma_{\ph\nu}^{\mu}\equiv\delta_{\nu}^{\mu}+u^{\mu}u_{\nu}\,,\label{eq:projectors-1+3}
\end{equation}
where all indices are manipulated with $g_{\mu\nu}$. It is easy to
check that $t_{\mu\nu}$ projects any tensor into the time direction,
whereas $\gamma_{\mu\nu}$ projects any tensor in the subspace orthogonal
to $u^{\mu}$. Moreover, one has
\[
t_{\ph\lambda}^{\mu}t_{\ph\nu}^{\lambda}=t_{\ph\nu}^{\mu}\,,\quad t_{\ph\mu}^{\mu}=1,\quad\gamma_{\ph\lambda}^{\mu}\gamma_{\ph\nu}^{\lambda}=\gamma_{\ph\nu}^{\mu}\,,\quad\gamma_{\ph\mu}^{\mu}=3\,,\qquad u^{\mu}\gamma_{\mu\nu}=0\,.
\]
Quite generally, the tensor $\gamma_{\mu\nu}$ acts as the metric
of local rest frames, since any two 3-vectors $A^{\mu}$ and $B^{\mu}$
will have a scalar product given by $g_{\mu\nu}A^{\mu}B^{\nu}=\gamma_{\mu\nu}A^{\mu}B^{\nu}$.
However, we are interested in the case where $u^{\mu}$ is everywhere
orthogonal to the spatial hypersurfaces\footnote{This is tantamount to assuming that Frobenius theorem ($u_{[\nu}\nabla_{\lambda}u_{\mu]}=0$)
holds.}, which further implies that $\gamma_{\mu\nu}$ is the metric on these
hypersurfaces (here called $\Sigma_{\tau}$ – see figure \ref{fig:splittings}). 

Given the above projectors, any 4-vector $v^{\mu}$ can be covariantly
decomposed into one scalar degree of freedom (d.o.f) in the time direction
plus a 3-vector orthogonal to it:
\begin{equation}
v^{\mu}=-Vu^{\mu}+V^{\mu}\,,\qquad u_{\mu}V^{\mu}=0\,,\label{eq:vmu-1+3}
\end{equation}
where $V\equiv u_{\mu}v^{\mu}$ and $V^{\mu}\equiv\gamma_{\ph\nu}^{\mu}v^{\nu}$.
Likewise, any (symmetric) rank-two tensor $h_{\mu\nu}$ can be uniquely
decomposed into two scalars $A$ and $B$ (1 d.o.f each), one 3-vector
$C_{\mu}$ (3 d.o.f) and one traceless 3-tensor $H_{\mu\nu}$ (5 d.o.f)
as

\begin{equation}
h_{\mu\nu}=2Au_{\mu}u_{\nu}+B\gamma_{\mu\nu}+2C_{(\mu}u_{\nu)}+H_{\mu\nu}\label{eq:tmunu-1+3}
\end{equation}
where, by definition
\begin{equation}
u^{\mu}C_{\mu}=0=u^{\mu}H_{\mu\nu}\,,\qquad\gamma^{\mu\nu}H_{\mu\nu}=0\,.\label{eq:D-properties}
\end{equation}
In terms of its irreducible components, then, the 10 d.o.f of $h_{\mu\nu}$
are split as 1+1+3+5 components in the 1+3 splitting. Conversely,
scalars, 3-vectors and trace-free 3-tensors can be extracted from
$h_{\mu\nu}$ in the following manner
\begin{align}
\begin{aligned}A & =u^{\alpha}u^{\beta}h_{\alpha\beta}/2\,,\qquad\\
B & =\gamma^{\alpha\beta}h_{\alpha\beta}/3\,,
\end{aligned}
 & \begin{aligned}C_{\mu} & =-u^{\alpha}\gamma_{\mu}^{\ph\beta}h_{\alpha\beta}\,,\\
H_{\mu\nu} & =\left(\gamma_{\mu}^{\ph\alpha}\gamma_{\nu}^{\ph\beta}-\frac{1}{3}\gamma_{\mu\nu}\gamma^{\alpha\beta}\right)h_{\alpha\beta}\equiv h_{\left\langle \alpha\beta\right\rangle }\,
\end{aligned}
\label{eq:1+3-irred-pieces}
\end{align}
Note that these equations implicitly define mode-extraction operators
which can be applied to $h_{\mu\nu}$ to obtain a specific mode. In
particular, the last equations defines the projection operator
\begin{equation}
P_{\mu\ph\nu}^{\ph\alpha\ph\beta}\equiv\left(\gamma_{\mu}^{\ph\alpha}\gamma_{\nu}^{\ph\beta}-\frac{1}{3}\gamma_{\mu\nu}\gamma^{\alpha\beta}\right)\,,\label{eq:operator-p}
\end{equation}
which extracts the traceless tensor component of $h_{\mu\nu}$. This
operator will be crucial to obtain the propagating degrees of freedom
of gravitational waves in the following section.

\subsubsection{Kinematics}

The existence of the projectors (\ref{eq:projectors-1+3}) allows
us to define two important tensorial derivatives of tensor fields:
\begin{equation}
\dot{q}_{\ph\nu_{1}\cdots}^{\mu_{1}\cdots}\equiv u^{\lambda}\nabla_{\lambda}q_{\ph\nu_{1}\cdots}^{\mu_{1}\cdots}\,,\quad\text{and}\quad D_{\rho}q_{\ph\nu_{1}\cdots}^{\mu_{1}\cdots}\equiv\gamma_{\rho}^{\ph\lambda}\gamma_{\ph\alpha_{1}}^{\mu_{1}}\cdots\gamma_{\nu_{1}}^{\ph\beta_{1}}\cdots\nabla_{\lambda}q_{\ph\beta_{1}\cdots}^{\alpha_{1}\cdots}\,,\label{eq:derivs}
\end{equation}
where $q_{\ph\nu_{1}\cdots}^{\mu_{1}\cdots}$ is an arbitrary \emph{spacetime}
tensor. These derivatives measure variations along and orthogonally
to $u^{\mu}$, respectively. Note in particular that, for any scalar
field $\varphi$, one has $\nabla_{\mu}\varphi=-u_{\mu}\dot{\varphi}+D_{\mu}\varphi$. 

A central kinematical quantity in the 1+3 splitting is the covariant
derivative of the fundamental observer's four velocity:
\begin{equation}
\nabla_{\mu}u_{\nu}\equiv-u_{\mu}A_{\nu}+K_{\mu\nu}\,,\label{eq:u-deriv}
\end{equation}
where $A_{\mu}\equiv\dot{u}_{\mu}$ is the observer's acceleration
and $K_{\mu\nu}\equiv D_{\mu}u_{\nu}$ is the extrinsic curvature
of spatial hypersurfaces. The latter measures spatial deformations
of timelike curves, which can be divided into an expansion ($K_{\ph\mu}^{\mu}$),
shear ($K_{\left\langle \mu\nu\right\rangle }$) and vorticity ($K_{[\mu\nu]}$)
of the congruence. Here we are interested in geodesic observers in
shear-free and irrotational universes, so that from now on we set
\begin{equation}
A_{\mu}=K_{\left\langle \mu\nu\right\rangle }=K_{[\mu\nu]}=0\,.\label{eq:shear-free-cond}
\end{equation}
In other words, $K_{\mu\nu}$ is a pure trace in the models we are
considering, measuring only the expansion of timelike congruences:
\[
K_{\mu\nu}\equiv\frac{\Theta}{3}\gamma_{\mu\nu}
\]
where $\Theta$ is a scalar measuring the global expansion of geodesics.
This reflects the fact that, from a kinematical point of view, shear-free
universes behave exactly like FLRW ones \cite{Carneiro:2001fz}. Finally,
we give for future reference two important relations that one can
easily check:
\[
\dot{\gamma}_{\mu\nu}=0=D_{\alpha}\gamma_{\mu\nu}\,.
\]

\subsection{1+2+1 spacetime splitting}

\begin{figure}
\begin{centering}
\includegraphics[scale=0.34]{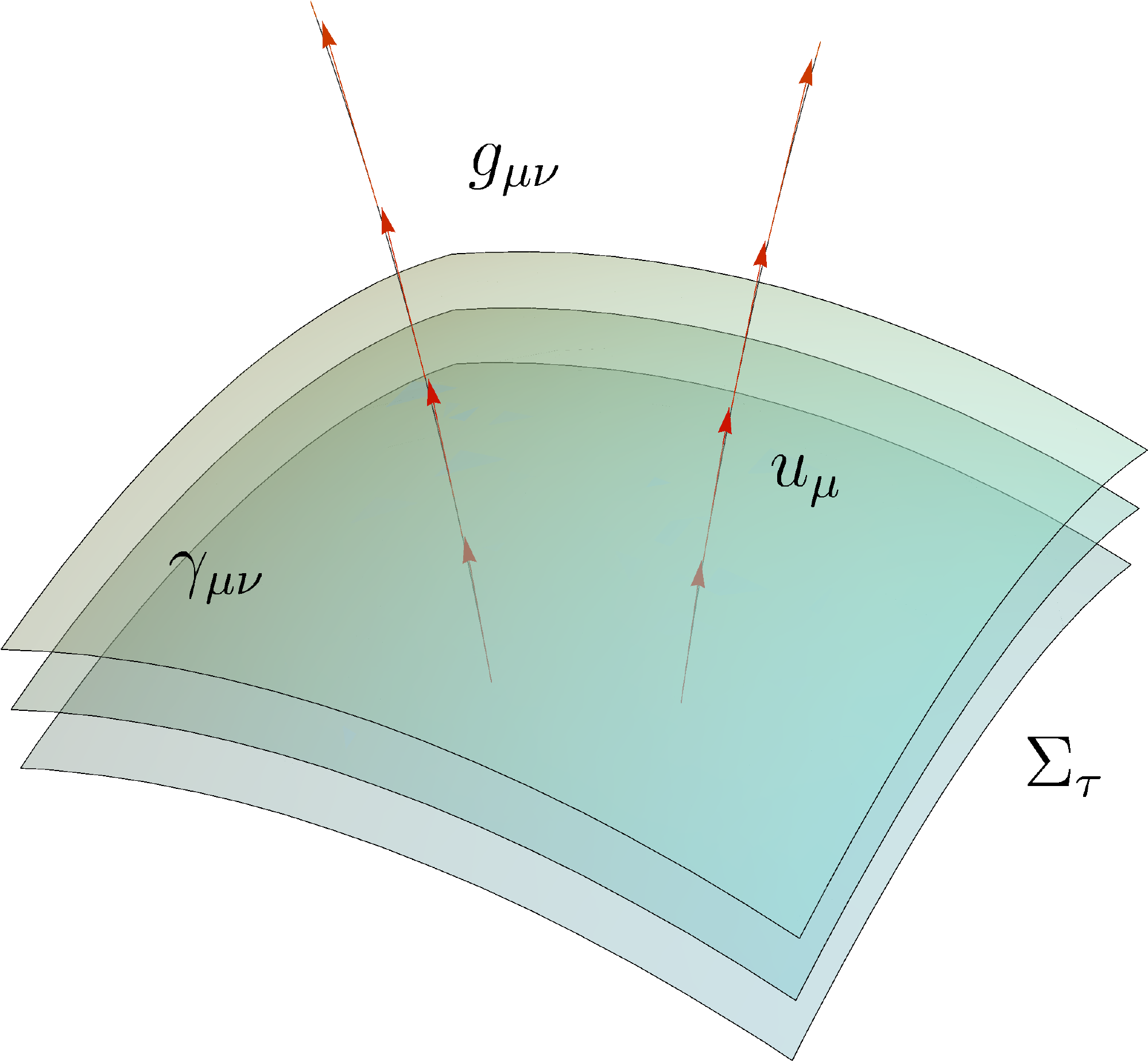}\includegraphics[scale=0.34]{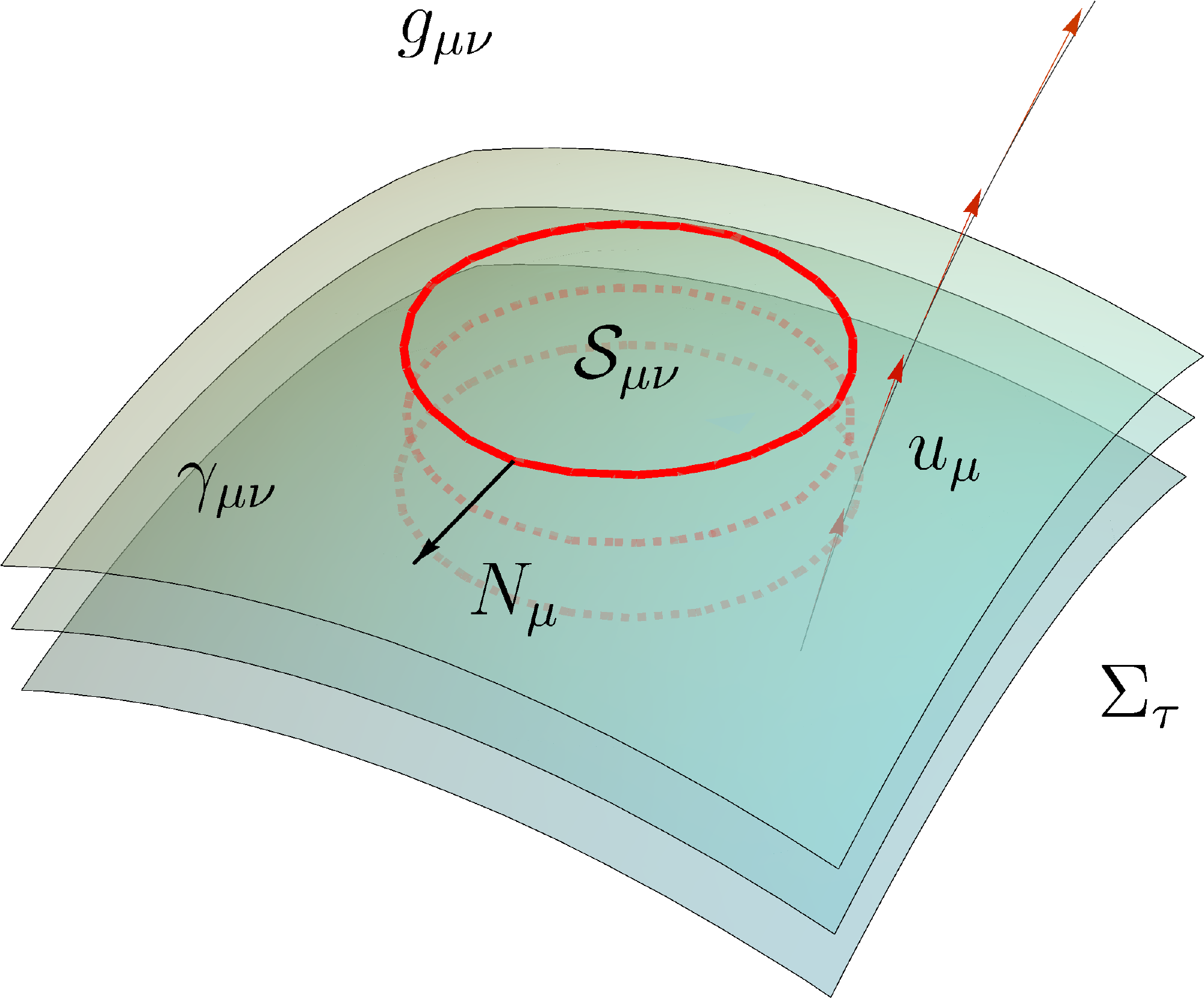}
\par\end{centering}
\caption{Schematic comparison between the 1+3 (left) and 1+2+1 (right) splittings.
$\Sigma_{\tau}$ labels the three-dimensional constant-time hypersurfaces.
The two-dimensional sheet ${\cal M}^{2}$ with metric $\mathcal{S}_{\mu\nu}$
is represented on the right panel by a closed (red) curve.}

\label{fig:splittings}
\end{figure}

The 1+3 spacetime splitting is quite general since it does not depend
on the specific symmetries of the constant time hypersurfaces $\Sigma_{\tau}$
(left panel of figure \ref{fig:splittings}). However, if $\Sigma_{\tau}$
has a privileged spacelike direction, which is the case of LRS (locally
rotationally symmetric) spacetimes with anisotropic curvature, then
we can covariantly split $\gamma_{\mu\nu}$ into components along
and orthogonal to this direction. Let $N^{\mu}$ be a unit spacelike
vector defining this direction. Such vector naturally defines two
projection tensors on $\Sigma_{\tau}$:
\begin{equation}
N_{\ph\nu}^{\mu}\equiv N^{\mu}N_{\nu}\,,\quad\text{and}\quad\mathcal{S}_{\ph\nu}^{\mu}\equiv\gamma_{\ph\nu}^{\mu}-N^{\mu}N_{\nu}\,,\label{eq:1+2+1-projectors}
\end{equation}
where $N_{\ph\nu}^{\mu}$ projects any tensor into the preferred spatial
direction and $\mathcal{S}_{\ph\nu}^{\mu}$ projects any tensor orthogonally
to it. From the above definitions it is easy to see that
\[
N_{\ph\lambda}^{\mu}N_{\ph\nu}^{\lambda}=N_{\ph\nu}^{\mu}\,,\quad N_{\ph\mu}^{\mu}=1,\quad\mathcal{S}_{\ph\lambda}^{\mu}\mathcal{S}_{\ph\nu}^{\lambda}=\mathcal{S}_{\ph\nu}^{\mu}\,,\quad\mathcal{S}_{\ph\mu}^{\mu}=2\,,\qquad\mathcal{S}_{\ph\nu}^{\mu}N^{\nu}=0\,.
\]
Being spatial projectors, they also satisfy
\begin{equation}
N_{\ph\nu}^{\mu}u^{\nu}=0=\mathcal{S}_{\ph\nu}^{\mu}u^{\nu}\,.
\end{equation}
In the particular case we are interested, where $\Sigma_{\tau}={\cal M}^{2}\times\mathbb{R}$,
$\mathcal{S}_{\mu\nu}$ defines the metric on the two-dimensional
subspace ${\cal M}^{2}$ and is everywhere orthogonal to $N^{\mu}$
— see the right panel on figure \ref{fig:splittings}. 

Given the vectors $u^{\mu}$ and $N^{\mu}$, and the metric $\mathcal{S}_{\mu\nu}$,
we would like to carry an irreducible decomposition of general tensors
in terms of scalars, 2-vectors, and traceless 2-tensors, where the
tracefree condition is now defined with respect to $\mathcal{S}_{\mu\nu}$.
We start by noting that any four-vector $v^{\mu}$ can be uniquely
written in terms of two scalars and one 2-vector as
\[
v^{\mu}=-Vu^{\mu}+\mathcal{U}N^{\mu}+\mathcal{V}^{\mu}\,,\qquad N_{\mu}\mathcal{V}^{\mu}=u_{\mu}\mathcal{V}^{\mu}=0\,,
\]
where $V\equiv u_{\mu}v^{\mu}$, $\mathcal{U}\equiv N_{\mu}v^{\mu}$
and $\mathcal{V}^{\mu}\equiv\mathcal{S}_{\ph\nu}^{\mu}v^{\nu}$. A
straightforward comparison with (\ref{eq:vmu-1+3}) reveals that $V^{\mu}=\gamma_{\ph\nu}^{\mu}v^{\nu}=\left(\mathcal{S}_{\ph\nu}^{\mu}+N^{\mu}N_{\nu}\right)v^{\nu}=\mathcal{V}^{\mu}+\mathcal{U}N^{\mu}$,
as it should be, since the new splitting only affects spatially-projected
quantities. 

Moving forward, we can split any symmetric and rank-two tensor uniquely
as
\begin{equation}
h_{\mu\nu}=2{\cal A}u_{\mu}u_{\nu}+{\cal B}\mathcal{S}_{\mu\nu}+2{\cal C}N_{\mu}N_{\nu}+2{\cal Q}N_{(\mu}u_{\nu)}+2\mathcal{E}_{(\mu}u_{\nu)}+2\mathcal{F}_{(\mu}N_{\nu)}+\mathcal{G}_{\mu\nu}\label{eq:tmunu-1+2+1}
\end{equation}
where, by force of our notation
\begin{equation}
u^{\mu}\mathcal{E}_{\mu}=N^{\mu}\mathcal{E}_{\mu}=u^{\mu}\mathcal{F}_{\mu}=N^{\mu}\mathcal{F}_{\mu}=0\,,\quad\text{and}\quad u^{\mu}\mathcal{G}_{\mu\nu}=N^{\mu}\mathcal{G}_{\mu\nu}=0=\mathcal{S}^{\mu\nu}\mathcal{G}_{\mu\nu}\,.\label{eq:transversality-1+2+1}
\end{equation}
Looking to the right hand side of (\ref{eq:transversality-1+2+1})
we would be tempted to conclude that $\mathcal{G}_{\mu\nu}$ has only
one independent d.o.f. However this conclusion is false since the
condition $N^{\mu}\mathcal{G}_{\mu\nu}=0$ eliminates only three-variables
once $u^{\mu}\mathcal{G}_{\mu\nu}=0$ is implemented\footnote{To see how this happens, suppose we adopt coordinates such that $u^{\mu}=\delta_{0}^{\mu}$
and $N^{\nu}=\delta_{\nu_{*}}^{\nu}$, for some fixed $\nu_{*}\neq0$.
Then $u^{\mu}\mathcal{G}_{\mu\nu}=0$ tell us that $\mathcal{G}_{0\nu*}=0$.
But since $\mathcal{G}_{\mu\nu}$ is symmetric, the equation $N^{\nu}\mathcal{G}_{\nu0}=0$
gives no new information. Since the latter is a tensorial condition,
this results holds in any coordinate system.}. Thus we conclude that $h_{\mu\nu}$ splits as $1+1+1+1+2+2+2$ irreducible
pieces in the 1+2+1 splitting. Contrarily, given $h_{\mu\nu},$ these
irreducible components can be extracted as follows
\begin{align}
\begin{aligned}{\cal A} & =u^{\alpha}u^{\beta}h_{\alpha\beta}/2\,,\qquad\\
{\cal B} & =\mathcal{S}^{\alpha\beta}h_{\alpha\beta}/2\,,\qquad\\
{\cal C} & =N^{\alpha\beta}h_{\alpha\beta}/2\,,\\
{\cal Q} & =-u^{\alpha}N^{\beta}h_{\alpha\beta}\,,
\end{aligned}
 & \begin{aligned}\mathcal{E}_{\mu} & =-u^{\alpha}\mathcal{S}_{\mu}^{\ph\beta}h_{\alpha\beta}\,,\qquad\\
\mathcal{F}_{\mu} & =N^{\alpha}\mathcal{S}_{\mu}^{\ph\beta}h_{\alpha\beta}\,,\\
\mathcal{G}_{\mu\nu} & =\left(\mathcal{S}_{\mu}^{\ph\alpha}\mathcal{S}_{\nu}^{\ph\beta}-\frac{1}{2}\mathcal{S}_{\mu\nu}\mathcal{S}^{\alpha\beta}\right)h_{\alpha\beta}\equiv h_{\{\alpha\beta\}}\,.\\
\\
\end{aligned}
\label{eq:1+2+1-irred-pieces}
\end{align}

\subsubsection{Kinematics}

The introduction of the projectors $N^{\mu}$ and $\mathcal{S}_{\mu\nu}$
allow us to define two new projected derivatives \cite{Clarkson:2002jz}:
\begin{equation}
q_{\ph\,\nu_{1}\cdots}^{\,\prime\mu_{1}\cdots}\equiv N^{\lambda}D_{\lambda}q_{\ph\nu_{1}\cdots}^{\mu_{1}\cdots},\quad{\cal D}_{\rho}q_{\ph\nu_{1}\cdots}^{\mu_{1}\cdots}\equiv\mathcal{S}_{\rho}^{\ph\lambda}\mathcal{S}_{\ph\alpha_{1}}^{\mu_{1}}\cdots\mathcal{S}_{\nu_{1}}^{\ph\beta_{1}}\cdots D_{\lambda}q_{\ph\beta_{1}\cdots}^{\alpha_{1}\cdots}\,,\label{eq:derivs2}
\end{equation}
where $q_{\ph\nu_{1}\cdots}^{\mu_{1}\cdots}$ is an arbitrary spacetime
tensor. It is important to note that these definitions are made with
respect to the spatially projected operator $D_{\mu}$, and not to
the full spacetime derivative $\mathbf{\nabla}_{\mu}$. Since $\mathcal{S}_{\mu\alpha}\gamma_{\ph\nu}^{\alpha}=\mathcal{S}_{\mu\nu}$,
this has no effect on the definition of the operator ${\cal D}_{\mu}$.
However, it \emph{does} change the definition of the derivative along
$N^{\mu}$ (here represented by a prime) since, for any spatial 3-vector
$V_{\mu}$, one has
\begin{align*}
V'_{\mu} & =N^{\lambda}D_{\lambda}V_{\mu}\\
 & =N^{\lambda}\left(\gamma_{\lambda}^{\ph\alpha}\gamma_{\mu}^{\ph\beta}\nabla_{\alpha}V_{\beta}\right)\\
 & =N^{\alpha}\nabla_{\alpha}V_{\mu}+u_{\mu}u^{\beta}\left(N^{\alpha}\nabla_{\alpha}V_{\beta}\right)\,.
\end{align*}
In other words, this definition ensures that $u^{\mu}V'_{\mu}=0$,
as one can easily check.

In analogy to (\ref{eq:u-deriv}), it will also be necessary to find
the irreducible decomposition of the projected tensor $D_{\mu}N_{\nu}$.
This is given by\footnote{In deriving this expression we have used $N^{\beta}D_{\alpha}N_{\beta}=-N^{\beta}D_{\alpha}N_{\beta}=0$.}
\[
D_{\mu}N_{\nu}=N_{\mu}{\cal A}_{\nu}+{\cal K}_{\mu\nu}\,,
\]
where $\mathcal{A}_{\nu}\equiv N'_{\nu}$ measures the observer's
acceleration along $N_{\nu}$, and $\mathcal{K}_{\mu\nu}\equiv\mathcal{D}_{\mu}N_{\nu}$
is the extrinsic curvature of ${\cal M}^{2}$. In analogy to $K_{\mu\nu}$,
it measures deformations of a congruence of curves along the privileged
direction. As before, such deformations can be separated into an expansion
($\mathcal{K}_{\ph\mu}^{\mu}$), a shear (\emph{$\mathcal{K}_{\{\mu\nu\}}$})
and a torsion ($\mathcal{K}_{[\mu\nu]}$) of the bundle. However,
it is important to stress that the quantities defined by $\mathcal{K}_{\mu\nu}$
are not in the exact same footing as those defined by $K_{\mu\nu}$,
since the former define deformations of spacelike curves on the same
surface $\Sigma_{\tau}$, whereas the latter defines deformation as
time evolves, thus connecting $\Sigma_{\tau}$ to $\Sigma_{\tau+{\rm d}\tau}$.
Moreover, we stress that the torsion term $\mathcal{K}_{[\mu\nu]}$
is identically zero by virtue of our choice of ${\cal S}_{\mu\nu}$
as the global metric on ${\cal M}^{2}$ (i.e., Frobenius theorem).
Furthermore, since BIII and KS spacetimes are spatially homogeneous,
we must have
\begin{equation}
\mathcal{K}_{\ph\mu}^{\mu}=0=\mathcal{K}_{\{\mu\nu\}}\,.\label{eq:spatial-deform}
\end{equation}
For the same reason, we should not experience any acceleration as
we travel along $N^{\mu}$, for this would imply a preferred position
in space. Thus we set
\begin{equation}
{\cal A}_{\mu}=0\,.\label{eq:z-deriv}
\end{equation}
Clearly, this conclusion would not be the same if we were working,
for example, in a Schwarzschild spacetime, where radial accelerations
do appear \cite{Clarkson:2002jz}. In conclusion, then, we have that
\begin{equation}
D_{\mu}N_{\nu}=0\label{eq:-dmunnu}
\end{equation}
for the BIII and KS models. Note however that this does not imply
that $\nabla_{\mu}N_{\nu}$ is zero. Indeed we have
\[
\nabla_{\mu}N_{\nu}=u_{\nu}N^{\alpha}K_{\mu\alpha}-u_{\mu}\dot{N}_{\nu}\,.
\]
At last, it follows from eqs. (\ref{eq:-dmunnu}) and (\ref{eq:shear-free-cond})
that
\[
\dot{{\cal S}}_{\mu\nu}+2\dot{N}_{(\mu}N_{\nu)}=0={\cal S}'_{\mu\nu}\,,\qquad D_{\alpha}S_{\mu\nu}=0={\cal D}_{\alpha}S_{\mu\nu}\,.
\]
These results will be used in the next section to identify the gravitational
waves d.o.f in such models.

\subsection{1+3 to 1+2+1 dictionary}

\begin{figure}
\begin{centering}
\includegraphics[scale=0.45]{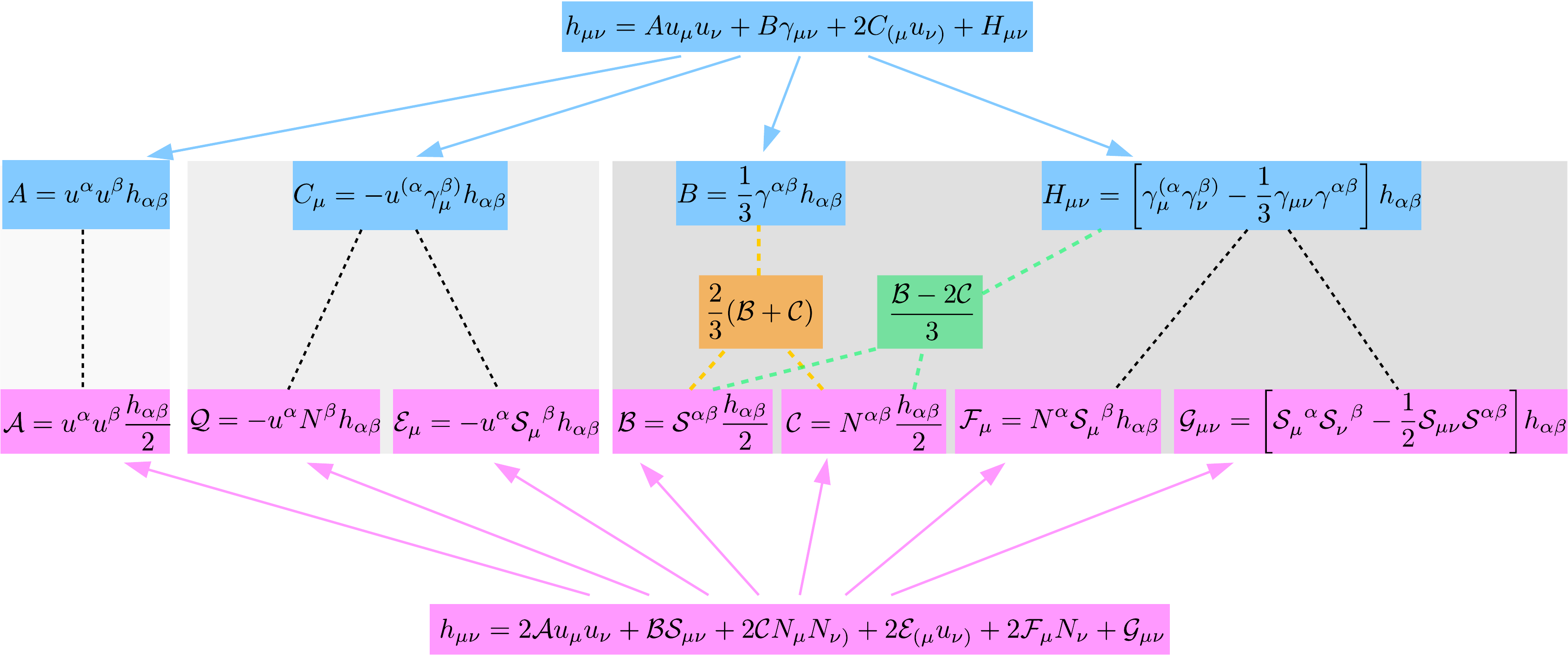}
\par\end{centering}
\caption{\textcolor{black}{Relation between the degrees of freedom of the 1+3
and the 1+2+1 splittings. Capital upright letters refer to 1+3 splitting,
while capital calligraphic letters refers to 1+2+1 spacetime splitting.
Note that the variables ${\cal B}$ and ${\cal C}$ contribute both
to a scalar mode ($B$) and to a transverse traceless mode $(H_{\mu\nu})$
of the 1+3 splitting. See the text for more details.}}

\label{fig:diagram}
\end{figure}

As a final task of this section we ask how the irreducible pieces
of the 1+2+1 decomposition add up to give the components of the 1+3
splitting, and vice-versa. The answer can be easily obtained by applying
the projectors implicitly defined in eqs. (\ref{eq:1+3-irred-pieces})
in the decomposition (\ref{eq:tmunu-1+2+1}). A straightforward computation
gives
\begin{equation}
\begin{aligned}A & =\mathcal{A}\,,\qquad\\
B & =\frac{2}{3}\left(\mathcal{B}+\mathcal{C}\right),\quad
\end{aligned}
\begin{aligned}C_{\mu} & =\mathcal{Q}N_{\mu}+\mathcal{E}_{\mu}\,,\\
H_{\mu\nu} & =\frac{\left({\cal B}-2{\cal C}\right)}{3}\left(\mathcal{S}_{\mu\nu}-2N_{\mu}N_{\nu}\right)+2\mathcal{F}_{(\mu}N_{\nu)}+\mathcal{G}_{\mu\nu}\,.
\end{aligned}
\label{eq:dictionary1}
\end{equation}
We can also obtain the inverse relations by applying the projectors
implicitly defined in (\ref{eq:1+2+1-irred-pieces}) into the tensor
(\ref{eq:tmunu-1+3}). This gives
\begin{align}
\begin{aligned}{\cal A} & =A\,,\qquad\\
{\cal B} & =B+\mathcal{S}^{\alpha\beta}H_{\alpha\beta}/2\,,\quad\\
{\cal C} & =\left(B+N^{\alpha\beta}H_{\alpha\beta}\right)/2\,,\\
{\cal Q} & =N^{\alpha}C_{\alpha}\,.
\end{aligned}
 & \begin{aligned}\mathcal{E}_{\mu} & =\mathcal{S}_{\mu}^{\ph\alpha}C_{\alpha}\,,\quad\qquad\mathcal{G}_{\mu\nu}=\left[\mathcal{S}_{\mu}^{\ph\alpha}\mathcal{S}_{\nu}^{\ph\beta}-\frac{1}{2}\mathcal{S}_{\mu\nu}\mathcal{S}^{\alpha\beta}\right]H_{\alpha\beta}\,,\\
\mathcal{F}_{\mu} & =N^{\alpha}\mathcal{S}_{\mu}^{\ph\beta}H_{\alpha\beta}\,,\\
\\
\\
\end{aligned}
\label{eq:dictionary2}
\end{align}
These relations are summarized in the diagram of figure \ref{fig:diagram}.
 We call the reader's attention to the algebraic mode coupling arising
between the 1+2+1 scalars ${\cal B}$ and ${\cal C}$ and the 1+3
quantities $B$ and $H_{\mu\nu}$. Indeed, from (\ref{eq:dictionary1})
we see that two independent combinations of ${\cal B}$ and ${\cal C}$
will contribute to both $B$ and $H_{\mu\nu}$, which are quantities
of rather different nature in the 1+3 splitting. We will come back
to this issue in the next section since, as we shall see, it is central
to our considerations.

We conclude this section by stressing that the above results are completely
general, and can be applied to any spacetime whose topology of spatial
sections are of the form $\Sigma_{\tau}={\cal M}^{2}\times\mathbb{R}$.
For example, in ref. \cite{Clarkson:2002jz} the 1+2+1 splitting was
used to investigate the evolution of gravitational waves in Schwarzschild
spacetimes, where $\Sigma_{\tau}=\mathbb{S}^{2}\times\mathbb{R}$.
Incidentally, note that the Schoolchild spacetime has the same spatial
topology as the Kantowski-Sachs universe which we consider here. The
main difference between these two spacetimes is in their set of isometries:
the former describes an static and inhomogeneous spacetime, whereas
the latter represents an expanding spatially homogeneous universe.

\section{Gravitational waves\label{sec:Gravitational-waves}}

We now move on to one of our main tasks, which is the identification
of the genuine tensor degrees of freedom in spacetimes with anisotropic
curvature. As a warm up exercise — and also to elucidate the difficulties
of this task — we briefly recall how this is done in FLRW universes
in \S\ref{subsec:FLRW-case}, after which we move to the more challenging
cases of BIII and KS spacetimes in \S\ref{subsec:BIII-and-KS}. Before
that, though, we explore the conformally static character of the family
of metrics (\ref{bckgd-metric}) to simplify some of our calculations.
More details can be found in Appendix \S\ref{appendix:conformal}.

\subsection{Conformal transformation}

Both spacetimes considered in this work are conformally related to
a static (background) metric — see (\ref{bckgd-metric}). In defining
metric perturbations, it is wise to separate perturbations of the
dynamic sector from those of the static one. We thus define 
\begin{equation}
\delta g_{\mu\nu}\equiv h_{\mu\nu}\,,\qquad\text{and}\qquad\delta\widetilde{g}_{\mu\nu}\equiv a^{2}h_{\mu\nu}\,,\label{eq:delta-g}
\end{equation}
where $h_{\mu\nu}$ is given by either (\ref{eq:tmunu-1+3}) or (\ref{eq:tmunu-1+2+1}).
Under a coordinate transformation of the form
\begin{equation}
x^{\mu}\rightarrow x^{\mu}-\xi^{\mu}\,,\label{eq:gauge-transf}
\end{equation}
where $\xi^{\mu}$ is the (infinitesimal) gauge vector\footnote{Note that we do not include a minus sign in the time component of
$\xi^{\mu}$ so as to comply with the standard literature, where $\xi^{\mu}=\left(T,L^{i}\right)$
in a comoving frame \cite{Mukhanov:1990me,Stewart:1990fm}.}, 
\[
\xi^{\mu}\equiv u^{\mu}T+J^{\mu}
\]
each sector will transform as (see Appendix \S\ref{appendix:conformal})
\begin{align}
\delta g_{\mu\nu} & \rightarrow\delta g_{\mu\nu}+2\nabla_{(\mu}\xi_{\nu)}\,,\label{eq:lie-gauge}\\
\delta\widetilde{g}_{\mu\nu} & \rightarrow\delta\widetilde{g}_{\mu\nu}+a^{2}\left(2\nabla_{(\mu}\xi_{\nu)}+2g_{\mu\nu}{\cal H}T\right)\,.\label{eq:lie-gauge-conf}
\end{align}
In the static sector many kinematical quantities are exactly zero,
which makes the computation of (\ref{eq:lie-gauge}) straightforward.
The effect of the expansion can be later included by using (\ref{eq:lie-gauge-conf}).
Indeed, in the static sector we have
\begin{equation}
K_{\mu\nu}=0=\dot{N}_{\mu}\label{eq:Kmunu-static}
\end{equation}
which greatly simplifies the computation of gauge transformations.
Once this is achieved, we can convert back to the dynamical sector
by using (\ref{eq:lie-gauge-conf}) and the fact that (see \S\ref{appendix:conformal})
\begin{align*}
\widetilde{K}_{\mu\nu} & =a\left(K_{\mu\nu}+{\cal H}\gamma_{\mu\nu}\right)\,,\\
\widetilde{\nabla}_{\mu}\widetilde{N}_{\nu} & =a\left(\nabla_{\mu}N_{\nu}+{\cal H}N_{\mu}u_{\nu}\right)\,.
\end{align*}

\subsection{FLRW case\label{subsec:FLRW-case}}

In FLRW universes, the constant-time hypersurface $\Sigma_{\tau}$
describes a maximally symmetric submanifold and, as we have seen,
the 1+3 spacetime splitting allow us to uniquely identify the tensor
degrees of freedom of metric perturbations, $\delta g_{\mu\nu}$,
with a traceless 3-tensor, such as $H_{\mu\nu}$ in (\ref{eq:tmunu-1+3}).
Note however that $H_{\mu\nu}$ has in general five degrees of freedom,
three more than the two polarization states of gravitational waves.
Such discrepancy, as is well known, is due to the fact that $H_{\mu\nu}$
is composed of fields of spin 0, 1 and 2, of which only the latter
corresponds to true gravitational waves, while the other two correspond
to pure gauge modes. Indeed, for a maximally symmetric subspace $\Sigma_{\tau}$,
$H_{\mu\nu}$ can be uniquely decomposed as 

\begin{equation}
H_{\mu\nu}=\left(D_{\mu}D_{\nu}-\frac{1}{3}\gamma_{\mu\nu}\nabla^{2}\right)2E+2D_{(\mu}E_{\nu)}+E_{\mu\nu}\,,\label{eq:svt}
\end{equation}
where
\[
\nabla^{2}\equiv D^{\mu}D_{\mu}\,,\qquad D^{\mu}E_{\mu}=0=D^{\mu}E_{\mu\nu}\,.
\]
This is the covariant formulation of the scalar-vector-tensor (SVT)
decomposition, and is unique up to boundary conditions at infinity
\cite{Stewart:1990fm}. This decomposition splits the five components
of $H_{\mu\nu}$ into one scalar $E$ (1 d.o.f), one divergence-free
3-vector $E_{\mu}$ (2 d.o.f) and one divergence-free and traceless
3-tensor $E_{\mu\nu}$ (2 d.o.f). The fact that both $E$ and $E_{\mu}$
are pure gauge modes can be seen by finding the 1+3 decomposition
of the transformation (\ref{eq:lie-gauge}). Using the conditions
(\ref{eq:shear-free-cond}), plus the fact that $K_{\mu\nu}=0$ in
the static sector of the metric, it is straightforward to show that
\begin{equation}
\delta g_{\mu\nu}\rightarrow\delta g_{\mu\nu}-2u_{\mu}u_{\nu}\dot{T}+2u_{(\mu}D_{\nu)}T+2D_{(\mu}J_{\nu)}-2u_{(\mu}\dot{J}_{\nu)}\,.\label{eq:hmunu-grauge-transf}
\end{equation}
We now project both sides of this expression with $P_{\mu\ph\nu}^{\ph\alpha\ph\beta}$
(see eq. (\ref{eq:operator-p})) to find
\begin{equation}
H_{\mu\nu}\rightarrow H_{\mu\nu}+2D_{(\mu}J_{\nu)}-\frac{2}{3}\gamma_{\mu\nu}D^{\alpha}J_{\alpha}\,.\label{eq:Hmunu-transf}
\end{equation}
Next, we carry an SVT decomposition of the vector $J_{\alpha}$, 
\[
J_{\alpha}=D_{\alpha}L+L_{\alpha}\,,\quad\text{where}\quad D^{\alpha}L_{\alpha}=0\,,
\]
and compare the result with (\ref{eq:svt}). This gives
\begin{align}
E & \rightarrow E+L\,,\nonumber \\
E_{\mu} & \rightarrow E_{\mu}+L_{\mu}\,,\label{eq:vector-gauge-transf}\\
E_{\mu\nu} & \rightarrow E_{\mu\nu}\,.\nonumber 
\end{align}
Finally, we need to change back to the dynamical perturbations $\delta\widetilde{g}_{\mu\nu}$
using (\ref{eq:lie-gauge-conf}). But since this is tantamount to
adding a trace to (\ref{eq:hmunu-grauge-transf}), which goes away
when projecting with $P_{\mu\ph\nu}^{\ph\alpha\ph\beta}$, we conclude
that the transformations (\ref{eq:vector-gauge-transf}) hold in general.
Clearly, $E$ and $E_{\mu}$ can be eliminated by going to a gauge
where $L=-E$ and $L_{\mu}=-E_{\mu}$, whereas $E_{\mu\nu}$ remains
gauge invariant. The 2 d.o.f in $E_{\mu\nu}$ represents the two polarizations
of gravitational waves, and their equations of motion can be found
by linearizing Einstein equations about the background metric. At
this point it is convenient to adopt a comoving coordinate system
given by
\[
u_{\mu}=-\delta_{\mu}^{0}\,,\qquad\gamma_{\mu\nu}=\text{diag}\left(0,\gamma_{ij}\right)
\]
and fix the gauge by choosing $E=0=E_{\mu}$, so that the most general
line element for tensor perturbations about (\ref{bckgd-metric})
reads
\begin{align*}
{\rm d}s^{2} & =a^{2}\left(g_{\mu\nu}+H_{\mu\nu}\right){\rm d}x^{\mu}{\rm d}x^{\nu}\,,\\
 & =a^{2}\left[-{\rm d}\eta^{2}+(\gamma_{ij}+E_{ij}){\rm d}x^{i}{\rm d}x^{j}\right]\,.
\end{align*}
Note that we are allowed to focus on the evolution of $E_{\mu\nu}$
alone, since the other components of $\delta g_{\mu\nu}$ will not
couple (either algebraically or dynamically) at first order. After
a lengthy but well-known computation one then finds \cite{Mukhanov:1990me}
\begin{equation}
\ddot{E}{}_{ij}+2{\cal H}\dot{E}{}_{ij}-\nabla^{2}E_{ij}=0\,,\label{eq:gw-eq-flrw}
\end{equation}
where, we remind the reader, dots represent derivatives with respect
to conformal time. 

Besides being damped by the expansion of the universe, the most remarkable
feature of the equation above is that it holds for both polarization
states. This is a direct consequence of the maximal symmetry of the
background space, and, as we will see, no longer holds in the presence
of anisotropic spatial curvature. For further reference, we note that
in a radiation dominated universe where the scale factor evolves as
$a=a_{0}\eta$, equation (\ref{eq:gw-eq-flrw}) has a regular solution
at $\eta=0$ which in Fourier space is given by
\begin{equation}
E_{ij}(q,\eta)=C_{ij}\frac{\sqrt{q\eta}}{a(\eta)}J_{1/2}(q\eta)\,,\label{eq:Elambda}
\end{equation}
where $q$ is the Fourier wavenumber.

\subsection{BIII and KS cases\label{subsec:BIII-and-KS}}

The previous discussion relied on the identification of gravitational
waves with the spin-2 and gauge-invariant components of metric perturbations.
As we have seen, when $\Sigma_{\tau}$ is maximally symmetric we were
able to define a unique component with these features (which we called
$E_{\mu\nu}$), rendering the identification straightforward. If we
tried to extend this program to the case $\Sigma_{\tau}=\mathcal{M}^{2}\times\mathbb{R}$,
a naive guess would be to identify gravitational waves with $\mathcal{G}_{\mu\nu}$
since, as we have seen, it has 2 d.o.f and is the only traceless\emph{
}tensor related to $H_{\mu\nu}$ — which is known to contain gravitational
waves d.o.f. There are however two serious problems with this approach.
The first is that, in spacetimes with one privileged direction, the
SVT decomposition (\ref{eq:svt}) is no longer appropriate, and one
has to resort instead to a scalar-vector (SV) decomposition which,
by construction, cannot accommodate a divergence-free and traceless
tensor like $E_{\mu\nu}$ \cite{Pereira:2012ma}. This means that
the two d.o.f originally contained in $E_{\mu\nu}$ have to be reallocated
into the irreducible pieces of the SV decomposition, but a priori
there is no way of guessing how to perform such task. Secondly, equations
(\ref{eq:dictionary1}) and (\ref{eq:dictionary2}) show that there
is an algebraic mode-coupling between 1+2+1 and 1+3 variables. This
tell us that we cannot focus on the evolution of a particular mode
and set the others to zero, as one does in deriving (\ref{eq:gw-eq-flrw}).
In fact, a similar situation happens in Bianchi I spacetimes: when
doing perturbation theory, one finds that vector perturbations have
no dynamics, and arise only as a constraint between scalars and tensors
\cite{Pereira:2007yy}. Thus, even though vector modes do not grow
as time evolves\footnote{If not sourced initially, as is usually the case.},
they cannot be set to zero from the beginning, since this would lead
to the wrong equations of motion for scalars and tensors. 

\textcolor{black}{In face of the above discussion, our approach here
will be as follows: first, we move to the static sector and split
(\ref{eq:lie-gauge}) in the 1+2+1 fashion, keeping all its degrees
of freedom. This will allow us to properly construct gauge-invariant
variables and to find their correct equations of motion. Given these
transformations we can use (\ref{eq:1+2+1-irred-pieces}) to extract
the transformation of the variables $(\mathcal{B}-\mathcal{C})$,
${\cal F}_{\mu}$ and ${\cal G}_{\mu\nu}$, which are directly related
to $H_{\mu\nu}$ — see (\ref{eq:dictionary2}). This will tell us
which of these behave as pure gauge modes, and which represent physical
perturbations. Surprisingly, we will find that ${\cal G}_{\mu\nu}$
can be completely gauged away, which tell us that our naive guess
aforementioned would be totally misplaced. }We thus start by splitting
(\ref{eq:lie-gauge}) into 1+2+1 irreducible pieces. Actually, since
(\ref{eq:hmunu-grauge-transf}) was already (1+3)-decomposed, all
we need to do is to split its spatial components in its 2+1 pieces.
To do that we write the gauge 3-vector $J_{\mu}$ as
\[
J_{\mu}=\mathcal{J}N_{\mu}+\mathcal{J}_{\mu}\,,
\]
where ${\cal J}_{\mu}N^{\mu}=0$. Next, we need to split the tensor
$D_{(\mu}J_{\nu)}$ into its 2+1 components. First, we note that for
any scalar field, and in particular for $\mathcal{J}$, we have $D_{\mu}\mathcal{J}=\mathcal{J}'N_{\mu}+\mathcal{D}_{\mu}\mathcal{J}$
(recall the definitions made in (\ref{eq:derivs2})). Furthermore,
we have 
\begin{align*}
D_{\mu}\mathcal{J}_{\nu} & =\gamma_{\mu}^{\ph\alpha}\gamma_{\nu}^{\ph\beta}D_{\alpha}\mathcal{J}_{\beta}\,,\\
 & =\mathcal{S}_{\mu}^{\ph\alpha}\mathcal{S}_{\nu}^{\ph\beta}D_{\alpha}\mathcal{J}_{\beta}+\mathcal{S}_{\mu}^{\ph\alpha}N_{\nu}N^{\beta}D_{\alpha}\mathcal{J}_{\beta}+N_{\mu}N^{\alpha}\gamma_{\nu}^{\ph\beta}D_{\alpha}\mathcal{J}_{\beta}\,,\\
 & =\mathcal{D}_{\mu}\mathcal{J}_{\nu}+N_{\mu}\mathcal{J}'_{\nu}\,,
\end{align*}
where we have used (\ref{eq:-dmunnu}). Combining all the terms, using
(\ref{eq:z-deriv}) and again (\ref{eq:-dmunnu}), we find that
\[
D_{\mu}J_{\nu}=\mathcal{J}'N_{\mu}N_{\nu}+N_{\nu}\mathcal{D}_{\mu}\mathcal{J}+\mathcal{D}_{\mu}\mathcal{J}_{\nu}+N_{\mu}\mathcal{J}'_{\nu}\,,
\]
and, in particular, that $D_{\alpha}J^{\alpha}=\mathcal{J}'+\mathcal{D}_{\alpha}\mathcal{J}^{\alpha}$.
These equalities can now be used to rewrite (\ref{eq:hmunu-grauge-transf})
as
\begin{align*}
\delta g_{\mu\nu} & \rightarrow\delta g_{\mu\nu}-2u_{\mu}u_{\nu}\dot{T}+2u_{(\mu}N_{\nu)}T'+2u_{(\mu}{\cal D}_{\nu)}T-2u_{(\mu}N_{\nu)}\dot{{\cal J}}-2u_{(\mu}\dot{{\cal J}}_{\nu)}\\
 & \qquad\quad\;+2\mathcal{J}'N_{\mu}N_{\nu}+2N_{(\mu}\mathcal{D}_{\nu)}\mathcal{J}+2\mathcal{D}_{(\mu}\mathcal{J}_{\nu)}+2N_{(\mu}\mathcal{J}'_{\nu)}\,.
\end{align*}
In deriving this transformation we have also used $\dot{N}_{\mu}=0$,
which is true in the static sector. Using $\delta g_{\mu\nu}=h_{\mu\nu}$
and (\ref{eq:1+2+1-irred-pieces}), the projection into 1+2+1 variables
then gives
\begin{equation}
\begin{alignedat}{2}{\cal A} & \rightarrow{\cal A}-\dot{T}\,, & \qquad{\cal E}_{\mu} & \rightarrow{\cal E}_{\mu}+{\cal D}_{\mu}T-\dot{{\cal J}}_{\mu}\,,\\
{\cal B} & \rightarrow{\cal B}+{\cal D}_{\alpha}{\cal J}^{\alpha}\,, & \qquad{\cal F}_{\mu} & \rightarrow{\cal F}_{\mu}+\mathcal{D}_{\mu}\mathcal{J}+{\cal J}'_{\mu}\,,\\
{\cal C} & \rightarrow{\cal C}+{\cal J}'\,, & \qquad{\cal G}_{\mu\nu} & \rightarrow{\cal G}_{\mu\nu}+2{\cal D}_{(\mu}{\cal J}_{\nu)}-{\cal S}_{\mu\nu}{\cal D}_{\alpha}{\cal J}^{\alpha}\,.\\
{\cal Q} & \rightarrow{\cal Q}+T'-\dot{{\cal J}}\,,
\end{alignedat}
\label{eq:static-gauge-transf}
\end{equation}
where we have also used ${\cal S}'_{\mu\nu}=0$ and $\dot{{\cal S}}_{\mu\nu}=0$,
the latter again being true for the static metric. At this point,
the next logical step would be to carry the SVT decomposition of $\mathcal{E}_{\mu}$,
$\mathcal{F}_{\mu}$, $\mathcal{J}_{\mu}$ and $\mathcal{G}_{\mu\nu}$.
However $\mathcal{G}_{\mu\nu}$ cannot be decomposed in the same way
as (\ref{eq:svt}) since, in two dimensions, there is no nontrivial
traceless and divergence-free 2-tensor\footnote{In two dimensions, a symmetric and traceless tensor has 2 components.
The divergence-free condition then eliminates two more, thus leading
to a null (i.e., trivial) tensor.}. Nonetheless, we can still carry a SV decomposition as follows
\begin{align*}
{\cal E}_{\mu} & ={\cal D}_{\mu}\mathcal{E}+\widehat{\mathcal{E}}_{\mu}\,,\qquad\,\,\mathcal{D}_{\mu}\widehat{\mathcal{E}}^{\mu}=0\,,\\
\mathcal{J}_{\mu} & =\mathcal{D}_{\mu}\mathcal{L}+\widehat{\mathcal{L}}_{\mu}\,,\qquad\,\mathcal{D}_{\mu}\widehat{\mathcal{L}}^{\mu}=0\,,\\
\mathcal{F}_{\mu} & =\mathcal{D}_{\mu}\mathcal{F}+\widehat{\mathcal{F}}_{\mu}\,,\qquad\mathcal{D}_{\mu}\widehat{\mathcal{F}}^{\mu}=0\,,\\
\mathcal{G}_{\mu\nu} & =2\Delta_{\mu\nu}\mathcal{G}+2\mathcal{D}_{(\mu}\widehat{\mathcal{G}}_{\nu)}\,,\qquad\mathcal{D}_{\mu}\widehat{\mathcal{G}}^{\mu}=0\,,
\end{align*}
where $\Delta_{\mu\nu}\equiv\left(\mathcal{D}_{\mu}\mathcal{D}_{\nu}-\frac{1}{2}\mathcal{S}_{\mu\nu}\Delta^{2}\right)$
and $\Delta^{2}\equiv{\cal D}_{\alpha}{\cal D}^{\alpha}$. Plugging
this decomposition into (\ref{eq:static-gauge-transf}) and converting
back to the dynamical perturbations $\delta\widetilde{g}_{\mu\nu}$,
we finally find
\begin{equation}
\begin{alignedat}{3}{\cal A} & \rightarrow{\cal A}-\dot{T}-\mathcal{H}T\,, & \qquad & \widehat{\mathcal{E}}_{\mu} & \rightarrow\widehat{\mathcal{E}}_{\mu}-\dot{\widehat{{\cal L}}}_{\mu}\,,\\
{\cal B} & \rightarrow{\cal B}+\Delta^{2}\mathcal{L}+2\mathcal{H}T\,, & \qquad & \widehat{\mathcal{F}}_{\mu} & \rightarrow\widehat{\mathcal{F}}_{\mu}+\widehat{{\cal L}}'_{\mu}\,,\\
{\cal C} & \rightarrow{\cal C}+{\cal J}'+\mathcal{H}T\,, & \qquad & \widehat{\mathcal{G}}_{\mu} & \rightarrow\widehat{\mathcal{G}}_{\mu}+\widehat{\mathcal{L}}_{\mu}\,.\\
{\cal Q} & \rightarrow{\cal Q}+T'-\dot{{\cal J}}\,,\\
{\cal E} & \rightarrow{\cal E}+T-\dot{{\cal L}}\,,\\
{\cal F} & \rightarrow{\cal F}+\mathcal{J}+{\cal L}'\,,\\
{\cal G} & \rightarrow{\cal G}+{\cal L}\,,
\end{alignedat}
\label{eq:dynamic-gauge-transf}
\end{equation}
These completes the task of finding the general gauge transformation
for 1+2+1 modes into scalar and vector components. From the above
we can form the following gauge invariant variables\footnote{From this point forward there is no need to maintain our special convention
for the labeling of variables, and so we arbitrarily chose letters
to define gauge-invariant variables.}:
\begin{equation}
\begin{alignedat}{2}\Phi & \equiv-{\cal A}-\frac{1}{a}\left[a\left(\mathcal{E}+\dot{\mathcal{G}}\right)\right]^{\mbox{\ensuremath{\boldsymbol{\cdot}}}}\,, & \;\Omega_{\mu} & \equiv\widehat{\mathcal{F}}_{\mu}-\widehat{\mathcal{G}}'_{\mu}\,,\\
2\Psi & \equiv\mathcal{B}-2\mathcal{H}\left(\mathcal{E}+\dot{\mathcal{G}}\right)-\Delta^{2}\mathcal{G}\,, & \Gamma_{\mu} & \equiv\dot{\widehat{{\cal G}}}_{\mu}+\widehat{\mathcal{E}}_{\mu}\,,\\
\Pi & \equiv\mathcal{Q}+\dot{\mathcal{F}}-\left(\mathcal{E}+2\dot{\mathcal{G}}\right)\,,\\
3X & \equiv\left(\mathcal{B}-2\mathcal{C}\right)+2\left(\mathcal{F}-\mathcal{G}'\right)'-\Delta^{2}\mathcal{G}\,.
\end{alignedat}
\label{eq:giv-1+2+1}
\end{equation}
By the Stewart-Walker lemma \cite{Stewart:1974uz}, we know that (linearized)
Einstein equations can be written in terms of these six variables.
In practice, though, it is easier to work in the gauge where
\begin{equation}
\mathcal{E}=\mathcal{F}=\mathcal{G}=0=\widehat{\mathcal{G}}_{\mu}\,,\label{eq:gauge-choice}
\end{equation}
since the final equations can be trivially converted back to gauge
invariant variables\footnote{This happens because, in the gauge (\ref{eq:gauge-choice}), metric
perturbations are equal to gauge-invariant variables. By the Stewart-Walker
lemma we know that the final equations can be made gauge-invariant.
Thus, when going to an arbitrary gauge only gauge-invariant variables
will remain, and these can be abstracted from the equations obtained
in the original gauge \cite{Mukhanov:1990me}.}. Note also that this choice completely fix the gauge. 

What about the gravitational waves d.o.f? From the last of eqs. (\ref{eq:dictionary1})
and the definitions (\ref{eq:giv-1+2+1}), we can easily write $H_{\mu\nu}$
as a sum of gauge-dependent variables plus gauge-invariant terms:
\begin{align*}
H_{\mu\nu} & =\frac{1}{3}\left[\Delta^{2}{\cal G}-2\left({\cal F}-{\cal G}'\right)'\right]\left(\mathcal{S}_{\mu\nu}-2N_{\mu}N_{\nu}\right)+2\left[{\cal D}_{(\mu}{\cal F}+\widehat{{\cal G}}'_{\mu)}\right]N_{\nu}+{\cal G}_{\mu\nu}\\
 & +X\left(\mathcal{S}_{\mu\nu}-2N_{\mu}N_{\nu}\right)+2\Omega_{(\mu}N_{\nu)}\,.
\end{align*}
This should be contrasted to (\ref{eq:svt}). Clearly, $X$ and $\Omega_{\mu}$
represent the two physical perturbations we were looking for, whereas
the remaining terms represent gauge (unphysical) modes. In the coordinate
system defined by (\ref{eq:gauge-choice}) we thus have
\begin{equation}
H_{\mu\nu}=X\left(\mathcal{S}_{\mu\nu}-2N_{\mu}N_{\nu}\right)+2\Omega_{(\mu}N_{\nu)}\,.\label{eq:gw-b3ks}
\end{equation}
It is interesting to note that the mode ${\Gamma_\mu}$ does not contribute to gravitational waves. 
Indeed, it is not hard to convince oneself that the pairs $(\Phi,\Psi)$ on one hand and $(\Pi,\Gamma_\mu)$ on 
the other correspond respectively to the two scalar and vector modes of standard (FLRW) perturbation theory. 
In fact, during inflation $\Pi$ and $\Gamma_\mu$ are dynamically suppressed 
while $\Psi$ and $\Phi$ become proportional to each other in the absence of anosotropic stress, 
just as in standard (FLRW) perturbation 
theory~\cite{Pereira:2012ma}. We have thus completed our first main task, which was to find the
physical variables describing gravitational waves in BIII and KS spacetimes.
It tell us that, when $\Sigma_{\tau}=\mathcal{M}^{2}\times\mathbb{R}$,
the original spin-2 components of $H_{\mu\nu}$ now behaves as a spin-0
($X$) and a spin-1 ($\Omega_{\mu}$) field. Qualitatively, this result
is easy to understand. Since $\mathbb{R}$ is one dimensional, it
only admits scalar propagating modes. Over $\mathcal{M}^{2}$ we can
have both scalar and vector modes, but as we have seen, the scalar
part turns out to be a pure gauge. Finally, note that there is no
reason a priori to expect that these fields have the same dynamics
and, as we shall see, they do not.

\section{Dynamical equations and solutions\label{sec:Dynamical-equations-and}}

Equation (\ref{eq:gw-b3ks}) is the first main result of this work.
Our second task is to find the equations of motion for $X$ and $\Omega_{\mu}$.
As stressed before, because of algebraic couplings between scalar
and vectors modes, these equations cannot be found by setting the
other perturbations to zero, and we have to work with the most general
metric perturbations. In the gauge (\ref{eq:gauge-choice}), this
is given by
\begin{align*}
\delta\widetilde{g}_{\mu\nu} & =a^{2}h_{\mu\nu}\\
 & =-2\Phi u_{\mu}u_{\nu}+2\Psi{\cal S}_{\mu\nu}+(2\Psi-3X)N_{\mu}N_{\nu}+2\Pi N_{(\mu}u_{\nu)}+2\Gamma_{(\mu}u_{\nu)}+2\Omega_{(\mu}N_{\nu)}\,.
\end{align*}
Adopting a coordinate system defined by 
\[
u_{\mu}=-\delta_{\mu}^{0}\,,\qquad\mathcal{S}_{\mu\nu}=\left(0,\mathcal{S}_{ab},0\right)\,,\qquad N_{\mu}=\delta_{\mu}^{z}\,,
\]
where indices $a$, $b$ run from 1 to 2, the corresponding line element
reads
\begin{align}
{\rm d}s^{2} & =a^{2}\left[-(1+2\Phi){\rm d}\eta^{2}-2\Gamma_{a}{\rm d}^{a}x{\rm d}\eta-2\Pi{\rm d}z{\rm d}\eta\right.\nonumber \\
 & \qquad\qquad\left.+\left(1+2\Psi\right)\mathcal{S}_{ab}{\rm d}x^{a}{\rm d}x^{b}+2\Omega_{a}{\rm d}x^{a}{\rm d}z+\left(1+2\Psi-3X\right){\rm d}z^{2}\right]\,.\label{eq:perturbed-ds2}
\end{align}
This is the most general line element for linear perturbations in
BIII and KS spacetimes, and it contains both scalar and vector perturbations.
The linearized components of the Einstein tensor are presented in
the Appendix \ref{appendix:deltaG}, together with the expressions
for the linearized energy momentum tensor of the anisotropic scalar
field, and we now focus on the construction of the equations of motion.
Let us start by finding the equation for $X$. From eq. (\ref{eq:deltaGab})
we see that, in the absence of anisotropic stress in the perturbations
of the perfect fluid, its scalar component obeys
\[
{\cal D}_{(a}{\cal D}_{b)}Z=0\,,\qquad\left(a\neq b\right)\,,
\]
where $Z\equiv2\Psi-3X+2\Phi$. Since this is the Killing equation,
we can write ${\cal D}_{a}Z$ as
\begin{equation}
{\cal D}_{a}Z=f_{i}(\eta,z)\zeta_{a}^{i}\,,\label{eq:Z-eq}
\end{equation}
where the $\zeta_{a}^{i}$s are the three Killing vectors on ${\cal M}^{2}$.
Given the explicit form of these vectors in a coordinate system, it
is easy to show that partial derivatives of $Z$ will only commute
if $f_{i}=0$, which then implies that $Z=0$ (see the appendix \ref{appendix:Killing-vectors}
for a proof). We thus conclude that
\begin{equation}
2\Psi+2\Phi=3X\,.\label{eq:X-Psi-Phi}
\end{equation}
Next, we subtract eq. (\ref{eq:deltaGzz}) from the trace of eq. (\ref{eq:deltaGab})
and use (\ref{eq:deltaGza}) to eliminate $\Pi$. Using (\ref{eq:X-Psi-Phi})
this finally leads to
\begin{equation}
\ddot{X}+2{\cal H}\dot{X}-(\nabla^{2}+2\kappa)X=0\,,\label{eq:X}
\end{equation}
which is the desired equation. 

Next, we look for a dynamical equation for $\Omega_{a}$. From the
vector component of eq. (\ref{eq:deltaGab}) we find ${\cal D}_{(a}Z_{b)}=0$,
where $Z_{a}\equiv\Omega'_{a}+\dot{\Gamma}_{a}+2{\cal H}\Gamma_{a}$
and $a\neq b$. Once more, this tell us that $Z_{a}=g_{i}(\eta,z)\zeta_{a}^{i}$
with arbitrary $g_{i}$. In the hyperbolic (BIII) case, the fact that
cosmological perturbations go to zero at infinity fixes $g_{i}$ to
zero, since there are no Killing vectors with this property \cite{Stewart:1990fm}.
In the spherical (KS) case there is no ``spatial infinity'' and
$Z_{a}$ remains non-unique. However, it is clear that $g_{i}\zeta_{a}^{i}$
does not represent cosmological perturbations, and there is no loss
of generality in again setting $g_{i}=0$. We thus use the constraint
\[
\dot{\Gamma}_{a}+2{\cal H}\Gamma_{a}=-\Omega'_{a}
\]
in eq. (\ref{eq:deltaGza}), which allow us to obtain the following
equation for the vector component:
\begin{equation}
\ddot{\Omega}_{a}+2{\cal H}\dot{\Omega}_{a}-\left(\nabla^{2}+\kappa\right)\Omega_{a}=0\,.\label{eq:Omega_a}
\end{equation}

Equations (\ref{eq:X}) and (\ref{eq:Omega_a}) summarize our quest
to find the dynamics of tensor perturbations in spacetimes with anisotropic
curvature, and it is interesting at this point to compare them against
eqs. (\ref{eq:gw-eq-flrw}). As we have anticipated, in the anisotropic
case each mode has its own dynamics, the difference being more prominent
at large cosmological scales, where the effect of the anisotropic
curvature is larger. At small scales ($\nabla^{2}\gg\kappa$) the
dynamics of the two modes become degenerate and we recover (\ref{eq:gw-eq-flrw}).
Note however that this is not the only difference between these equations
and the isotropic ones. In fact, the Laplacian in (\ref{eq:X}) and
(\ref{eq:Omega_a}) have different spectra which appear as a different
signatures in the power spectrum of each perturbation. We shall now
investigate these issues in detail.

\subsection{Spatial eigenfunctions\label{subsec:Spatial-eigenfunctions}}

In order to solve the dynamical equations and evaluate the power spectra
of gravitational waves we need to know how to perform Fourier analysis
in spaces with anisotropic curvature. This requires knowledge of the
eigenfunctions $\phi_{\mathbf{q}}$ of the Laplacian in these geometries,
which are solutions of the eigenvalue problem
\[
\nabla^{2}\phi_{\mathbf{q}}=-q^{2}\phi_{\mathbf{q}}\,,\qquad q\equiv\left|\mathbf{q}\right|\in\mathbb{R}^{+}\,.
\]
For the metrics codified in (\ref{eq:bckgd-metric cylindrical}) we
have
\[
\nabla^{2}=\kappa\left[\frac{1}{\sin\bar{\rho}}\partial_{\bar{\rho}}\left(\sin\bar{\rho}\,\partial_{\bar{\rho}}\right)+\frac{1}{\sin^{2}\bar{\rho}}\partial_{\varphi}^{2}+\partial_{\bar{z}}^{2}\right],
\]
where $\bar{\rho}\equiv\sqrt{\kappa}\rho$ and $\bar{z}\equiv\sqrt{\kappa}z$.
Note that for $\kappa=-\left|\kappa\right|$ this automatically gives
the Laplacian of the BIII model. The eigenfunctions are given by \cite{Pereira:2015pxa,Adamek:2010sg,BlancoPillado:2010uw}

\begin{equation}
\phi_{\mathbf{q}}\left(\mathbf{x}\right)=\begin{cases}
N_{\ell m}P_{-1/2+i\ell}^{m}\left(\cosh\bar{\rho}\right)e^{im\varphi}e^{ik\bar{z}}\,, & \left(\text{BIII}\right)\\
N_{\ell m}P_{\ell}^{m}\left(\cos\bar{\rho}\right)e^{im\varphi}e^{ik\bar{z}}\,, & \left(\text{KS}\right)
\end{cases}\label{eq:eigenfunc}
\end{equation}
where $k$ (not to be confused with the curvature parameter $\kappa$)
is real, $m$ is an integer and $\ell$ is either real (BIII) or integer
(KS), but always positive. These functions are orthonormal provided
that we fix $N_{\ell m}$ as
\[
N_{\ell m}=\frac{1}{2\pi}\begin{cases}
\sqrt{(-1)^{m}\ell\tanh\pi\ell\frac{\Gamma\left(i\ell-m+1/2\right)}{\Gamma\left(i\ell+m+1/2\right)}}\,, & \left(\text{BIII}\right)\\
\sqrt{\frac{2\ell+1}{2}\frac{\left(\ell-m\right)!}{\left(\ell+m\right)!}}\,. & \left(\text{KS}\right)
\end{cases}
\]
One can also check that in the limit $\rho\ll1$ and $\ell\gg1$,
$\phi_{\mathbf{q}}(\mathbf{x})$ becomes 
\[
\phi_{\mathbf{q}}(\mathbf{x})\approx\ell^{1/2}J_{m}(\ell\rho)\frac{e^{im\varphi}}{\sqrt{2\pi}}\frac{e^{ikz}}{\sqrt{2\pi}}
\]
which, not surprisingly, are the eigenfunctions of a flat FLRW universe
in cylindrical coordinates. Moreover, we notice that the eigenvalues
$q^{2}$ are $m$-independent (reflecting the residual rotational
symmetry of these geometries) and given by
\[
q^{2}=\frac{1}{R_{c}^{2}}\begin{cases}
\ell^{2}+k^{2}+\frac{1}{4}\,, & (\text{BIII},\,\ell\text{ real})\\
\left(\ell+1/2\right)^{2}+k^{2}-\frac{1}{4}\,, & (\text{KS},\,\ell\text{ integer})
\end{cases}
\]
where we have used (\ref{eq:Rc}). We thus see that, in both cases,
$q$ has a fundamental lower bound given by\footnote{In the open case, the largest mode has $\ell=k=0$. However, $\ell=0$
in the closed case gives a monopole, and is thus removed from the
spectrum. Hence $\ell-1=k=0$ in this case.}
\begin{equation}
qR_{c}\equiv q_{*}\geq\begin{cases}
(1/2)\,, & (\text{BIII})\\
\sqrt{2}\,. & (\text{KS})
\end{cases}\label{eq:qstar}
\end{equation}
As an aside, note that we can place observational bounds on $q$ using
recent CMB data. The latest limits on the FLRW spatial curvature set
by the \emph{Planck} team and using CMB data alone gives $\Omega_{\kappa0}=-0.005_{-0.017}^{+0.016}$
at 95\% of confidence level \cite{Ade:2015xua}. This translates into
\[
R_{c}\gtrsim6.7H_{0}^{-1}\quad\text{(BIII)}\qquad\text{and}\qquad R_{c}\gtrsim4.7H_{0}^{-1}\quad\text{(KS)}\,,
\]
or, equivalently, 
\[
qH_{0}^{-1}\gtrsim0.07\quad\text{(BIII)}\qquad\text{and}\qquad qH_{0}^{-1}\gtrsim0.3\quad\text{(KS)}\,.
\]
The fact that constraints on $R_{c}$ are weaker than those on $L_{c}$
(see eq. (\ref{eq:Rc})) is a direct consequence of the (assumed)
statistical anisotropy of the data. At large scales, where cosmic
variance dominates, the temperature multipolar coefficients $a_{\ell m}$
with fixed $\ell$ and different $m$ become correlated \cite{Pereira:2015pxa},
meaning that there is less constraining power in a given temperature
spectrum $C_{\ell}$ than in the case of isotropic models\footnote{Recall that, when isotropy holds, $C_{\ell}=(2\ell+1)^{-1}\sum_{m}\left|a_{\ell m}\right|^{2}$
is a sum of $2\ell+1$ independent numbers.}.\\

Given the eigenfunctions (\ref{eq:eigenfunc}), any scalar perturbation
$Q$ can be formally decomposed as
\[
Q(\eta,\mathbf{x})=\int{\rm d}\mu_{\mathbf{q}}\,Q_{\mathbf{q}}(\eta)\phi_{\mathbf{q}}(\mathbf{x})\,,
\]
with the inverse given by
\[
Q_{\mathbf{q}}(\eta)=\int\sqrt{\gamma}{\rm d}^{3}\mathbf{x}\,Q(\eta,\mathbf{x})\phi_{\mathbf{q}}^{*}(\mathbf{x})\,.
\]
Here, $\mu_{\mathbf{q}}$ formally represents the integration measure
defined by the topology of each space (see \cite{Pereira:2015pxa}
for the details). The decomposition of a transverse vector field $Q_{a}$
is essentially the same if we recognize that a transverse vector in
two-dimensions is fundamentally a scalar field, and as such it can
be written as
\[
Q_{a}=\epsilon_{ab}{\cal D}^{b}Q
\]
where $\epsilon_{ab}$ is the volume two-form on ${\cal M}^{2}$.
Since $\epsilon_{ab}$ is covariantly conserved, this automatically
ensures that ${\cal D}^{a}Q_{a}=0$. Thus, in practice we will always
be working with scalars, and from now on we drop the index $a$ in
any transverse vector. Before we proceed it is interesting to introduce
two rescaled variables
\[
x\equiv aX\,,\qquad\omega\equiv a\Omega\,,
\]
in terms of which the equations of motion simplify (in Fourier space)
to
\begin{equation}
\begin{split}\ddot{x}_{\mathbf{q}}+\left(q^{2}-\frac{\ddot{a}}{a}-2\kappa\right)x_{\mathbf{q}} & =0\,,\\
\ddot{\omega}_{\mathbf{q}}+\left(q^{2}-\frac{\ddot{a}}{a}-\kappa\right)\omega_{\mathbf{q}} & =0\,.
\end{split}
\label{eq:reduced-equations}
\end{equation}
Note that the frequency terms of these oscillators (i.e., the terms
inside parenthesis) are fully isotropic, since they do not depend
on $\mathbf{q}$, but only on its modulus. This is again a reflection
of the isotropic background expansion, but also of the fact that the
1+2+1 splitting is adapted to the symmetries of the background space,
so that no dynamical mode coupling arises. In particular, this implies
that the quantization of the perturbations $x$ and $\omega$ during
inflation will proceed along the same lines as the quantization of
free fields in curved FLRW spaces \cite{Mukhanov:2007zz}, and that
the power spectrum of each perturbation will only depend on $q$.
We promptly stress, however, that the \emph{total} power spectrum
(i.e., the one defined through the complete tensor mode (\ref{eq:gw-b3ks}))
will certainly be a function of the full vector $\mathbf{q}$, rather
than just $q$, since its definition requires that we fix the direction
of the anisotropic curvature to some angle in the sky. In other words,
we are still free to fix the orientation of $\mathbb{R}$ relative
to ${\cal M}^{2}$. This will further imply in an anisotropic angular
correlation function and in off-diagonal terms in, say, the CMB temperature
covariance matrix \cite{Pereira:2015pxa}. We postpone a detailed
analysis of these and other observables effects to a future work.

\subsection{Solutions and power spectra}

\begin{figure}
\begin{centering}
\includegraphics[scale=0.58]{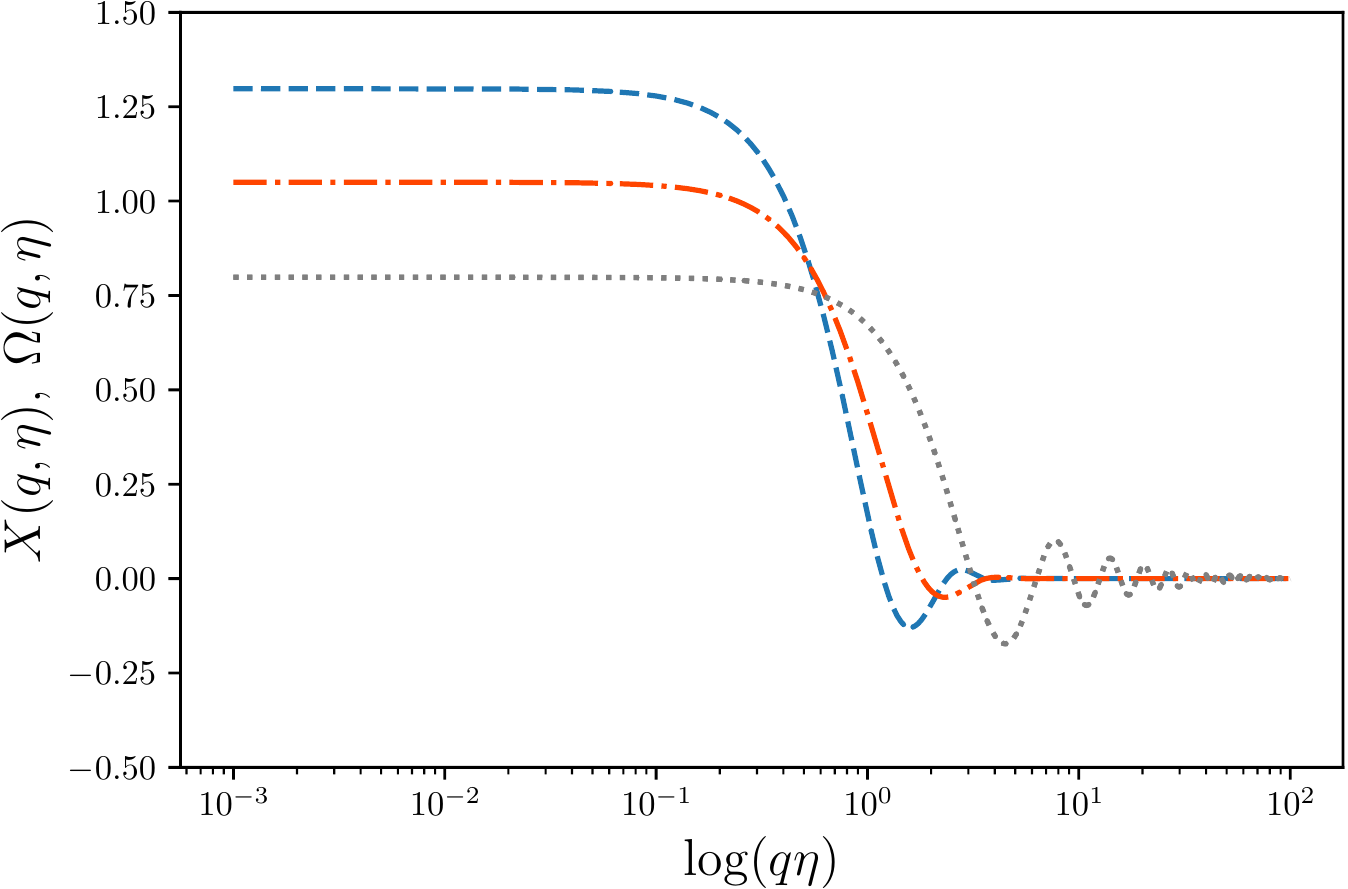}\includegraphics[scale=0.58]{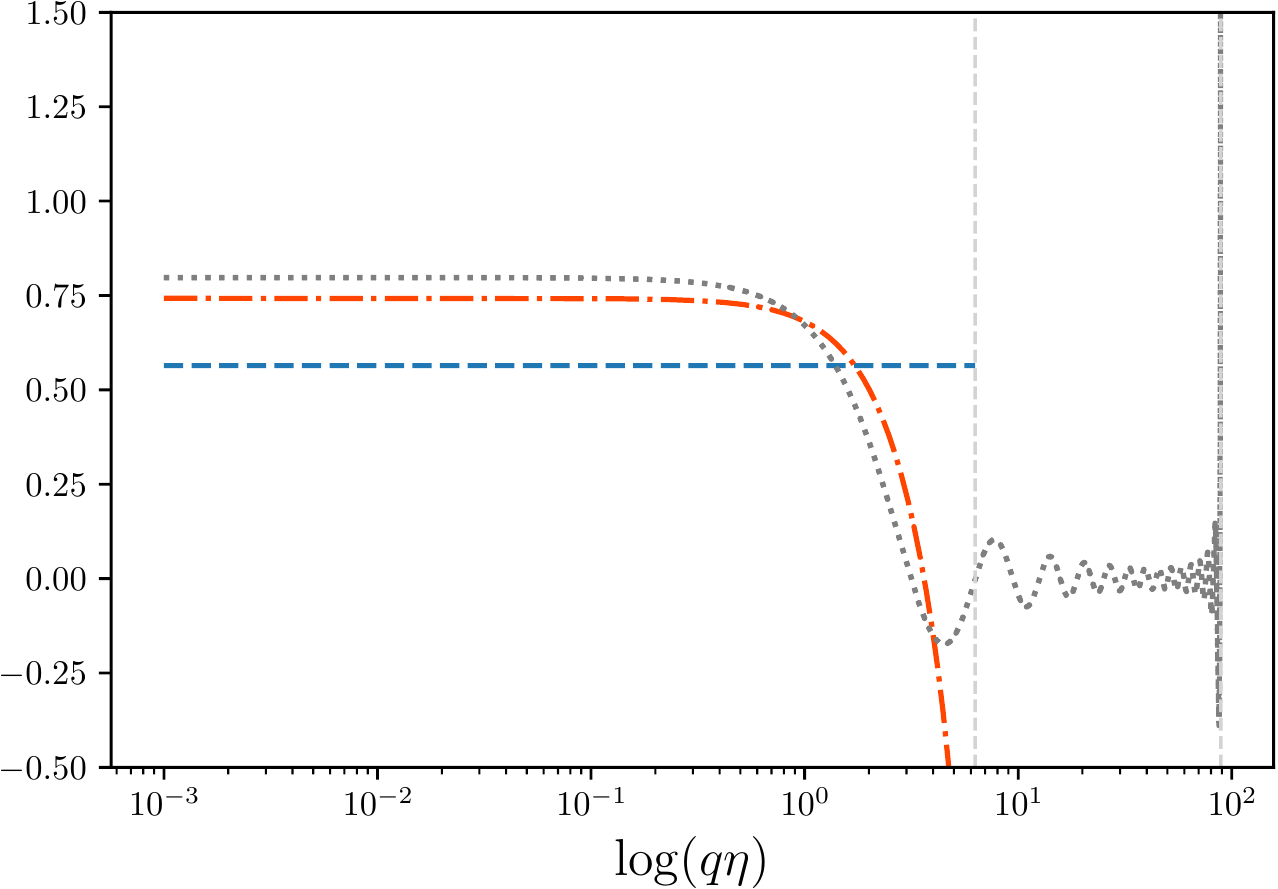}
\par\end{centering}
\caption{Evolution of the tensor modes $X$ (dashed line) and $\Omega$ (dot-dashed
line) in BIII (left) and KS (right) models corresponding to the longest
possible wavelength (see eq. (\ref{eq:qstar})). The plots also show
the evolution of the perturbations for $q_{*}=20$, in which case
$X$ and $\Omega$ match the isotropic evolution. The vertical lines
in the right panel represent the collapse time in the closed model
for different curvature radii. Note that $X(q_{*}=\sqrt{2})$ is constant
in the KS model for the whole evolution. See the text for details.
For these plots we have set $qC_{1,2}/a_{0}=1$.}

\label{fig:time-evolution}
\end{figure}

As a simple application of our results, let us find analytical solutions
to eqs. (\ref{eq:reduced-equations}) in some well known cosmological
regimes. The simplest and most important case is that of radiation
dominance for which $p_{f}=\rho_{f}/3$, for in that case equation
(\ref{eq:eff-flrw-eqs}) becomes simply
\[
\frac{\ddot{a}}{a}=-\frac{\kappa}{2}=-\frac{\epsilon}{2R_{c}^{2}}\,,\qquad\epsilon=\text{sign}\left(\kappa\right)\,.
\]
In this regime the scale factor evolves as
\[
a(\eta)=a_{0}\begin{cases}
\sqrt{2}R_{c}\sinh\left(\frac{\eta}{\sqrt{2}R_{c}}\right) & \left(\text{BIII, }\epsilon=-1\right)\,,\\
\sqrt{2}R_{c}\sin\left(\frac{\eta}{\sqrt{2}R_{c}}\right) & \left(\text{KS, }\epsilon=+1\right)\,,
\end{cases}
\]
where $a_{0}$ is an integration constant. Note that we have normalized
these solutions so that $a(\eta)=a_{0}\eta$ when $R_{c}\gg1$, which
is the correct solution in the isotropic case. If we define $\beta^{2}\equiv1-3\epsilon/2q_{*}^{2}$
and $\gamma^{2}\equiv1-\epsilon/2q_{*}^{2}$, the regular solutions
at $\eta=0$ are
\[
X(q,\eta)=C_{1}\frac{\sqrt{q\eta}}{a(\eta)}J_{1/2}\left(\beta q\eta\right),\quad\Omega(q,\eta)=C_{2}\frac{\sqrt{q\eta}}{a(\eta)}J_{1/2}\left(\gamma q\eta\right),
\]
where $C_{1,2}$ are integration constants. These solutions differ
from those at eq. (\ref{eq:Elambda}) in two aspects: first, the Bessel
functions now have an explicit dependence on the curvature radius
through the factors $\beta$ and $\gamma$, which in the isotropic
case are both equal to one. Second, the scale factor now has a completely
different behavior as the one of a flat FLRW radiation dominated universe.
One particularly interesting consequence of these solutions is that,
in a KS universe, the infinite wavelength perturbation of the tensor
mode $X$ is constant in time. This happens because, in KS spaces,
the largest wavelength has $q_{*}=\sqrt{2}$, which implies that $\beta=1/2$.
Since $J_{1/2}(u/2)=2\sin\left(u/2\right)/(\sqrt{\pi u})$, we immediately
find that $X=qC_{1}/a_{0}\sqrt{\pi}$ for all times. This is shown
in figure (\ref{fig:time-evolution}) together with some other solutions
for different values of the parameter $q_{*}$. For the sake of illustration
we have adopted $qC_{1,2}/a_{0}=1$ in these plots.

\begin{figure}
\begin{centering}
\includegraphics[scale=0.58]{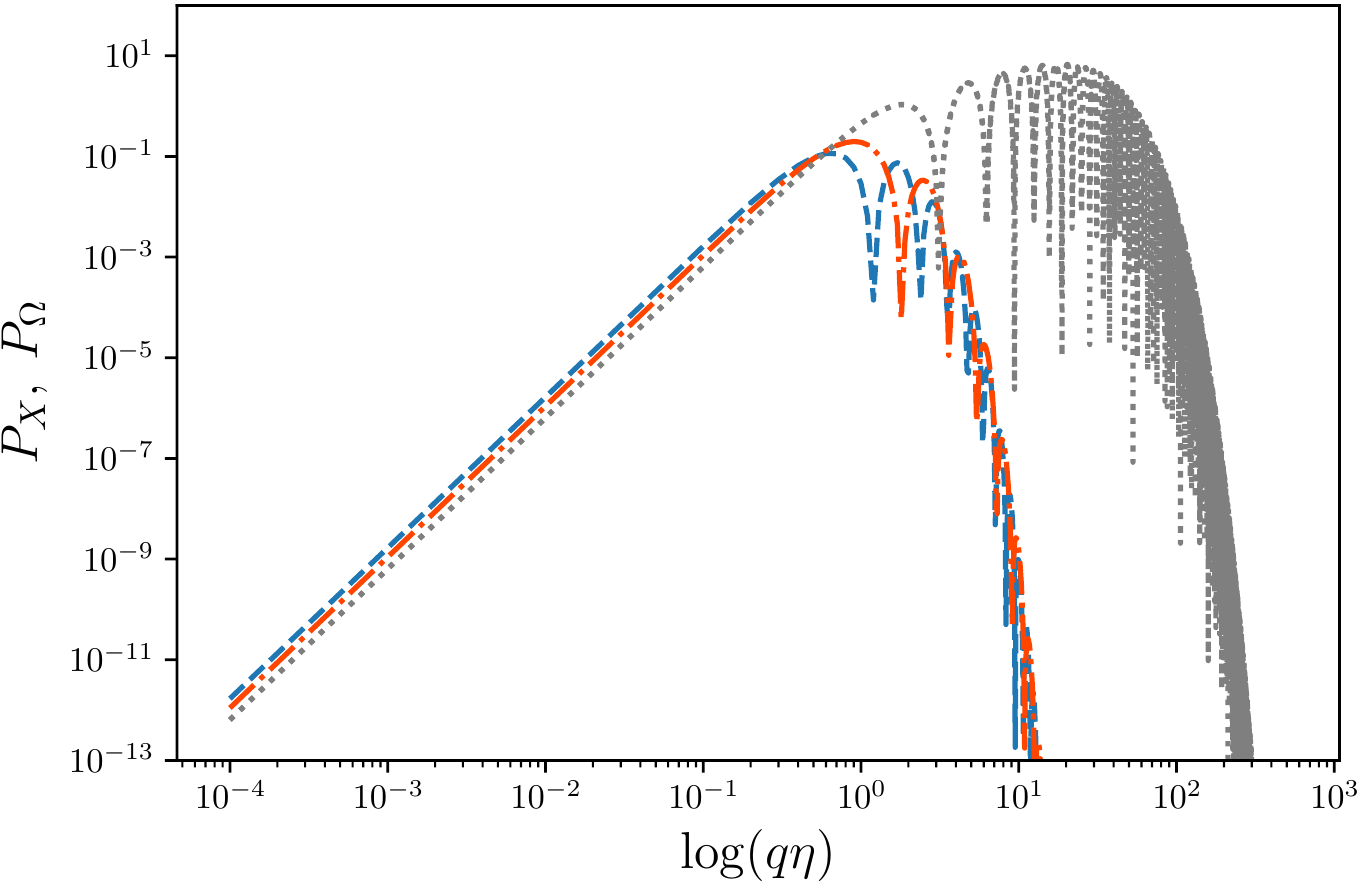}\includegraphics[scale=0.58]{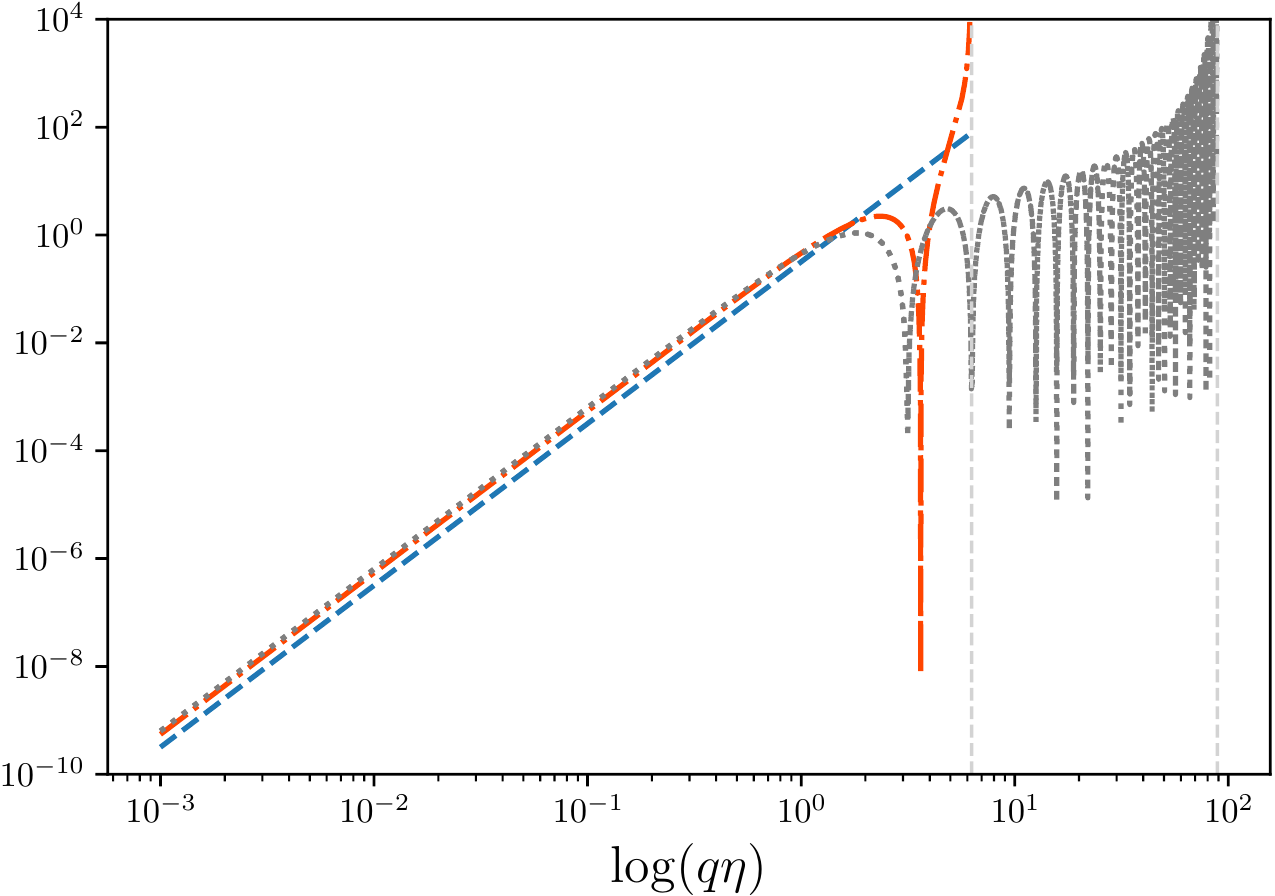}
\par\end{centering}
\caption{Power spectra $P_{X}$ and $P_{\Omega}$ of the tensor modes in BIII
(left) and KS (right) models, with the same parameters values and
coloring schemes as that of figure (\ref{fig:time-evolution}). Dashed
vertical lines on the right panel represent the collapse time for
each curvature radius. The amplitudes were arbitrarily set to one.}

\label{fig:psec-figure}
\end{figure}

Because the dynamics of the modes $X$ and $\Omega$ is isotropic,
we can define their power spectra simply as
\[
P_{X}=q^{3}\left|X\left(q,\eta\right)\right|^{2}\,,\qquad P_{\Omega}=q^{3}\left|\Omega(q,\eta)\right|^{2}\,.
\]
At large scales both $X$ and $\Omega$ are constant, and the power
spectrum scales as $q^{3}$. As the mode crosses the Hubble horizon
(i.e., at $q\eta\gtrsim1$) the power at each perturbation scales
nearly linearly with $q$. This goes on until the curvature affects
the behavior of the scale factor, in which case the power will either
decrease or increase, depending on whether the expansion goes on forever
or faces a recollapse. This is shown numerically in figure (\ref{fig:psec-figure}). 

\section{Conclusions and final remarks\label{sec:Conclusions}}

The detection of gravitational waves by the Ligo/Virgo collaboration
opens a plethora of new opportunities for astrophysics and cosmology.
While the direct detection of primordial gravitational waves might
yet take a while, the impact of tensor perturbations in the temperature
and polarization spectra of the CMB can already help us to constrain
models of the early universe. In this work we have explored a particular
class of models where the assumption of spatial isotropy is broken
in such a way that the CMB remains isotropic at the background level.
Such models offer an interesting example where the symmetries of the
universe do not follow from the symmetries of the data, as one usually
postulates. In particular, Bianchi type III and Kantowski-Sachs solutions
with an imperfect-fluid matter content lead to the same expansion
history as that of a curved Friedmann-Lemaître-Robertson-Walker universe.
Here we have paid particular attention to the theoretical construction
of linear gravitational waves in these models. In doing so, we found
that the main difficulty is to separate physical from gauge degrees
of freedom in the presence of anisotropic spatial curvature, which
requires the use of a mode-splitting well adapted to the symmetries
of the spacetime. Moreover, the presence of algebraic couplings between
modes prevents one from ignoring perturbations not related to gravitational
waves. In particular, had we kept only the $X$ and $\Omega_{a}$
variables in the expansion (\ref{eq:perturbed-ds2}) we would end
up with the wrong equations of motion.

Two specific signatures of gravitational waves arise in this context.
First, the polarization modes of the wave behave as a spin-0 and spin-1
irreducible components of a transverse and traceless tensor, rather
than two components of a spin-2 field. As such, each polarization
has its own dynamics which differs from the usual (isotropic) case;
the difference being larger at scales near the curvature radius. Second,
the presence of a curvature radius in these models naturally implies
in upper-limits to the size of a gravitational wave. In particular,
we have found that the largest wave corresponding to the $X$-polarization
mode in a KS universe is constant in time. However, such effect might
easily be hidden in the stochastic background of gravitational radiation,
and thus can be hard to detect. 

We would like to conclude by commenting on a few applications of the
formalism here presented. First, the integrated tensor Sachs-Wolfe
effect predicts a variation in temperature given by $\Delta T/T=-\int e^{i}e^{j}\partial_{\eta}E{}_{ij}{\rm d}\eta$
\cite{Peter:1208401}. Clearly, in the presence of anisotropic curvature
it will get different contributions from the $X$- and $\Omega$-polarization
modes, which will in turn affect the amplitude of the tensor spectrum
at large CMB angles. Moreover, because there is an upper-limit to
the maximum length of a wave, one should expect to find a deficit
in the tensor spectrum at large angles, similarly to what happens
with the tensor spectrum in closed FLRW universes \cite{Bonga:2016cje}.
Second, since the total power spectrum of tensor perturbations will
depend on the direction of the vector $N^{\mu}$, one might expect
anisotropies in the tensor/scalar ratio from inflation, which can
be used as a further signature to discriminate these models. Lastly,
one can also expect to find signatures coming from the direct effect
of gravitational waves in the polarization of the $B$-modes of CMB.
We postpone these analysis to a future work.

\pagebreak{}

\acknowledgments We would like to thank Pedro Gomes, Carlos Hernaski and Mikjel Thorsrud for enlightening remarks during the development of this work. We also thank Mikjel Thorsrud for a careful reading of the final version of this manuscript. This work was supported by Conselho Nacional de Desenvolvimento Tecnológico (under grant number 311732/2015-1) and Fundação Araucária (PBA 2016). FOF thanks CAPES for the financial support. 

\appendix
\section{Miscellanea}

We gather here some useful formulae and results which were used in
the main text.

\subsection{Conformal transformations}\label{appendix:conformal}

We give here some details about the expressions involving conformal
transformations. In the family of metrics in which we are interested,
the \emph{dynamic} (background) metric $\widetilde{g}_{\mu\nu}$ is
conformally related to the \emph{static} metric $g_{\mu\nu}$ through
\[
\widetilde{g}_{\mu\nu}(x^{\lambda})=a^{2}g_{\mu\nu}(x^{\lambda})\,,\qquad\widetilde{g}^{\mu\nu}(x^{\lambda})=a^{-2}g^{\mu\nu}(x^{\lambda})\,,
\]
where $a=a(\eta)$ is the scale factor and $x^{\lambda}=\left(\eta,\mathbf{x}\right)$.
The proper time one-forms associated to each metric are related to
each other through ${\rm d}\widetilde{\tau}=a{\rm d}\tau$. This further
implies that
\[
\widetilde{u}^{\mu}=\frac{dx^{\mu}}{d\widetilde{\tau}}=\frac{1}{a}\frac{dx^{\mu}}{d\tau}=a^{-1}u^{\mu}\,.
\]
The following relations then follow from consistency
\[
\widetilde{u}_{\mu}=au_{\mu}\,,\quad\widetilde{\gamma}_{\mu\nu}=a^{2}\gamma_{\mu\nu}\,,\quad\widetilde{N}_{\mu}=aN_{\mu}\,,\quad\widetilde{{\cal S}}_{\mu\nu}=a^{2}{\cal S}_{\mu\nu}\,.
\]
However, note that $\widetilde{\gamma}_{\mu}^{\ph\nu}=\gamma_{\mu}^{\ph\nu}$
and $\widetilde{{\cal S}}_{\mu}^{\ph\nu}={\cal S}_{\mu}^{\ph\nu}$.
Because both metrics describe homogeneous spacetimes, $\widetilde{\gamma}_{\mu}^{\ph\alpha}\widetilde{\nabla}_{\alpha}a=\widetilde{D}_{\mu}a=0$,
which implies that
\begin{equation}
\widetilde{\nabla}_{\mu}a=-\widetilde{u}_{\mu}\widetilde{u}^{\alpha}\widetilde{\nabla}_{\alpha}a=-u_{\mu}u^{\alpha}\nabla_{\alpha}a=\nabla_{\mu}a\,.\label{eq:u-tilde-deriv}
\end{equation}
In particular, in the comoving coordinate system of metric (\ref{bckgd-metric}),
$u^{\mu}=\left(1,0,0,0\right)$ and we have
\[
\nabla_{\mu}a=-u_{\mu}\dot{a}\,,\quad\text{where}\quad\dot{a}=\frac{da}{d\eta}\,.
\]
Given the gauge transformation of $\delta g_{\mu\nu}$, we can find
that of $\delta\widetilde{g}_{\mu\nu}$ as follows:
\begin{align*}
\delta\widetilde{g}_{\mu\nu} & \rightarrow\delta\widetilde{g}_{\mu\nu}+{\cal L}_{\xi}\left(\widetilde{g}_{\mu\nu}\right)\,,\\
 & =\delta\widetilde{g}_{\mu\nu}+a^{2}\left({\cal L}_{\xi}g_{\mu\nu}+2g_{\mu\nu}\frac{1}{a}\xi^{\alpha}\nabla_{\alpha}a\right)\,.
\end{align*}
By writing $\xi^{\mu}=u^{\mu}T+J^{\mu}$ and using (\ref{eq:u-tilde-deriv}),
it then follows that
\[
\delta\widetilde{g}_{\mu\nu}\rightarrow\delta\widetilde{g}_{\mu\nu}+a^{2}\left({\cal L}_{\xi}g_{\mu\nu}+2g_{\mu\nu}\mathcal{H}T\right)\,,
\]
where $\mathcal{H}=\dot{a}/a$ is the conformal Hubble function. 

The derivatives $\widetilde{\nabla}_{\mu}$ and $\nabla_{\mu}$ will
in general be different when acting on tensors, since
\begin{align*}
\widetilde{\Gamma}_{\mu\nu}^{\alpha} & =\Gamma_{\mu\nu}^{\alpha}+\frac{1}{a}\left(2\delta_{(\mu}^{\alpha}\nabla_{\nu)}a-g_{\mu\nu}\nabla^{\alpha}a\right)\,,\\
 & =\Gamma_{\mu\nu}^{\alpha}-\left(2\delta_{(\mu}^{\alpha}u_{\nu)}-g_{\mu\nu}u^{\alpha}\right)\mathcal{H}\,.
\end{align*}
In particular, this implies that the extrinsic curvature of the dynamic
and static metric will be related by
\[
\widetilde{K}_{\mu\nu}=\widetilde{\nabla}_{\mu}\widetilde{u}_{\nu}=a\left(K_{\mu\nu}+\mathcal{H}\gamma_{\mu\nu}\right)\,.
\]
Likewise, the covariant derivative of vector $\widetilde{N}_{\mu}$
is given by
\[
\widetilde{\nabla}_{\mu}\widetilde{N}_{\nu}=a\left(\nabla_{\mu}N_{\nu}+\mathcal{H}N_{\mu}u_{\nu}\right)\,.
\]
\subsection{Killing vectors on ${\cal M}^2$}\label{appendix:Killing-vectors}

Maximally symmetric two-dimensional manifolds of constant curvature
can be represented by the following line element:
\[
{\rm d}s^{2}={\rm d}\rho^{2}+S_{\kappa}^{2}(\rho){\rm d}\varphi^{2}
\]
where $S_{\kappa}(\rho)=\sin\left(\sqrt{\kappa}\rho\right)/\sqrt{\kappa}$.
These spaces admit three Killing vectors:
\begin{equation}
\begin{split}\boldsymbol{\zeta}^{(1)} & =\partial_{\varphi}\,,\\
\boldsymbol{\zeta}^{(2)} & =\cos\varphi\partial_{\rho}-\sqrt{\kappa}\cot\left(\sqrt{\kappa}\rho\right)\sin\varphi\partial_{\varphi}\,,\\
\boldsymbol{\zeta}^{(3)} & =\sin\varphi\partial_{\rho}+\sqrt{\kappa}\cot\left(\sqrt{\kappa}\rho\right)\cos\varphi\partial_{\varphi}\,.
\end{split}
\label{eq:killing-vectors}
\end{equation}
Referring back to eq. (\ref{eq:Z-eq}) we see that
\[
\partial_{\rho}Z=f_{2}\cos\varphi+f_{3}\sin\varphi\,,\qquad\partial_{\varphi}Z=f_{1}+\left(-f_{2}\sin\varphi+f_{3}\cos\varphi\right)\sqrt{\kappa}\cot\left(\sqrt{\kappa}\rho\right)\,.
\]
Clearly, the equality $\partial_{\varphi}\partial_{\rho}Z=\partial_{\rho}\partial_{\varphi}Z$
will hold for arbitrary $\rho$ and $\varphi$ only when $f_{2}=f_{3}=0$.
Moreover, since cosmological perturbations are defined as fluctuations
\emph{above} the mean, they don't have a monopole. But $f_{1}$ is
clearly a monopole, since it does not depend on $\rho$ and $\varphi$.
Thus $f_{1}=0$. We thus conclude that $\partial_{\rho}Z=0=\partial_{\varphi}Z$,
which implies that $Z$ is a constant. But since $Z$ cannot have
a monopole, we conclude that $Z=0$.

\subsection{Perturbed Einstein tensor}\label{appendix:deltaG}

We give here a brief overview of linear perturbation theory. More
details can be found in ref. \cite{Pereira:2012ma}. As usual, we
write the metric and its inverse as
\[
g_{\mu\nu}\rightarrow g_{\mu\nu}+\delta g_{\mu\nu}\,,\qquad g^{\mu\nu}\rightarrow g^{\mu\nu}+\delta g^{\mu\nu}\,,
\]
so that
\[
\delta g^{\mu\nu}=-g^{\mu\alpha}g^{\nu\beta}\delta g_{\alpha\beta}\,.
\]
In the coordinate system (\ref{eq:bckgd-metric cylindrical}) we have
\[
g_{00}=-a^{2}\,,\quad g_{ab}=a^{2}{\cal S}_{ab}\,,\quad g_{zz}=a^{2}
\]
with inverse
\[
g^{00}=-a^{-2}\,,\quad g^{ab}=a^{-2}{\cal S}^{ab}\,,\quad g^{zz}=a^{-2}\,.
\]
Metric perturbations are parameterized as
\[
\begin{split}\delta g_{00} & =-2a^{2}\Phi\,,\\
\delta g_{0a} & =-a^{2}\Gamma_{a}\,,\\
\delta g_{0z} & =-a^{2}\Pi\,,\\
\delta g_{ab} & =2a^{2}{\cal S}_{ab}\Psi\,,\\
\delta g_{az} & =a^{2}\Omega_{a}\,,\\
\delta g_{zz} & =a^{2}(2\Psi-3X)\,.
\end{split}
\]
These can be used to linearize all the tensors forming Einstein equations.
The final result is
\begin{align}
a^{2}\delta G_{\ph0}^{0} & =6{\cal H}^{2}\Phi-2{\cal H}\Pi'-3{\cal H}(2\dot{\Psi}-\dot{X})+\frac{1}{2}{\cal D}^{c}{\cal D}_{c}(4\Psi-3X)+2\kappa\Psi+2\Psi''\nonumber \\
 & =a^{2}\delta T_{\ph0}^{0}\,,\label{eq:deltaG00}\\
a^{2}\delta G_{\ph a}^{0} & ={\cal D}_{b}{\cal D}_{[a}\Gamma^{b]}-\frac{1}{2}\Gamma''_{a}-\kappa\Gamma_{a}-\frac{1}{2}\dot{\Omega}'_{a}+\frac{1}{2}{\cal D}_{a}\Pi'+\frac{1}{2}{\cal D}_{a}(4\dot{\Psi}-3\dot{X})-2{\cal H}{\cal D}_{a}\Phi\nonumber \\
 & =a^{2}\delta T_{\ph a}^{0}\,,\label{eq:deltaG0a}\\
a^{2}\delta G_{\ph z}^{0} & =-2{\cal H}\Phi'+2\dot{\Psi}'-\frac{1}{2}{\cal D}^{a}{\cal D}_{a}\Pi=a^{2}\delta T_{\ph z}^{0}\,,\label{eq:deltaG0z}\\
a^{2}\delta G_{\ph b}^{a} & =\delta_{b}^{a}\left[2\Phi\left({\cal H}^{2}+2\dot{{\cal H}}\right)+2{\cal H}\dot{\Phi}+\Psi''-2{\cal H}\Pi'-\frac{1}{2}(4\ddot{\Psi}-3\ddot{X})-{\cal H}(4\dot{\Psi}-3\dot{X})\right.\nonumber \\
 & \left.+\Phi''-\dot{\Pi}'+\frac{1}{2}{\cal D}^{c}{\cal D}_{c}(2\Psi-3X+2\Phi)\right]-\frac{1}{2}{\cal D}^{(a}{\cal D}_{b)}\left(2\Psi-3X+2\Phi\right)\nonumber \\
 & +{\cal D}^{(a}\Omega'_{b)}+{\cal D}^{(a}\dot{\Gamma}_{b)}+2{\cal H}{\cal D}^{(a}\Gamma_{b)}=a^{2}\delta T_{\ph b}^{a}\,,\label{eq:deltaGab}\\
a^{2}\delta G_{\ph a}^{z} & =-{\cal D}_{a}\left(\Phi'+\Psi'\right)+{\cal H}{\cal D}_{a}\Pi+\frac{1}{2}{\cal D}_{a}\dot{\Pi}+{\cal H}\Gamma'_{a}+\frac{1}{2}\dot{\Gamma}'_{a}+\frac{1}{2}\ddot{\Omega}_{a}+{\cal H}\dot{\Omega}_{a}-\frac{\kappa}{2}\Omega_{a}-\frac{1}{2}{\cal D}^{c}{\cal D}_{c}\Omega_{a}\nonumber \\
 & =a^{2}\delta T_{\ph a}^{z}\,,\label{eq:deltaGza}\\
a^{2}\delta G_{\ph z}^{z} & =2\Phi({\cal H}^{2}+2\dot{{\cal H}})+2{\cal H}\dot{\Phi}+{\cal D}^{c}{\cal D}_{c}\left(\Phi+\Psi\right)-2\left(\ddot{\Psi}+2{\cal H}\dot{\Psi}-\kappa\Psi\right)\nonumber \\
 & =a^{2}\delta T_{\ph z}^{z}\,.\label{eq:deltaGzz}
\end{align}
We remind the reader that a dot and a prime means $\partial_{\eta}$
and $\partial_{z}$, respectively. 

The non-zero perturbed components of the energy-momentum tensor are
\begin{align*}
a^{2}\delta T_{\ph0}^{0} & =\kappa\left(-\Psi+3X/2+\delta\phi'\right)\,,\\
a^{2}\delta T_{\ph z}^{0} & =\kappa\left(\Pi+\delta\phi'\right)\,,\\
a^{2}\delta T_{\ph b}^{a} & =\kappa\left(-\Psi+3X/2+\delta\phi'\right)\delta_{b}^{a}\,,\\
a^{2}\delta T_{\ph a}^{z} & =-\kappa\,{\cal D}_{a}\delta\phi\,,\\
a^{2}\delta T_{\ph z}^{z} & =\kappa\left(\Psi-3X/2-\delta\phi'\right)\,.
\end{align*}
Note that we are not perturbing the other components of the energy-momentum
tensor. This is equivalent to the assumption that the perturbed anisotropic
stress of the remaining fluids are negligible, which is a good approximation
at large scales. \pagebreak{}

\bibliographystyle{JHEP}
\bibliography{b3ks_signature}

\end{document}